\documentclass[iop]{emulateapj}
\usepackage{graphicx,color,natbib,enumerate}

\newcommand{\msun}{\,\rm M_\odot}
\newcommand{\Zsun}{\rm Z_\odot}
\newcommand{\Zsn}{\rm Z_{SN}}

\newcommand{\f}{\frac}

\newcommand{\HH}{H$_2$}
\newcommand{\fHH}{f_{{\rm H}_2}}
\newcommand{\SigmaSFR}{\Sigma_{\rm SFR}}
\newcommand{\SigmaGas}{\Sigma_{\rm gas}}
\newcommand{\SigmaHH}{\Sigma_{{\rm H}_2}}
\newcommand{\SigmaHI}{\Sigma_{\rm HI}}
\newcommand{\SigmaSob}{\Sigma_{\rm Sob}}
\newcommand{\SigmaCell}{\Sigma_{\rm cell}}
\newcommand{\rhoGas}{\rho_{\rm gas}}
\newcommand{\fstar}{\rm f_\star} 
\newcommand{\Mhalo}{\rm M_h}
\newcommand{\Mgas}{\rm M_{gas}}
\newcommand{\Mdm}{\rm M_{DM}}
\newcommand{\Mstar}{\rm M_\star}
\newcommand{\Zfloor}{\rm Z_{floor}}

\begin{document}

\submitted{}

\title{Dwarf galaxy formation with \HH-regulated star formation}

\author{Michael Kuhlen} 
\affil{Theoretical Astrophysics Center, University of 
California, Berkeley, CA 94720}
\email{mqk@astro.berkeley.edu}

\author{Mark R. Krumholz} 
\affil{Department of Astronomy \& Astrophysics,  University of 
California, Santa Cruz, CA 95064}

\author{Piero Madau} 
\affil{Department of Astronomy \& Astrophysics,  University of 
California, Santa Cruz, CA 95064}

\author{Britton D. Smith} 
\affil{Department of Physics and Astronomy, Michigan State University, East Lansing, MI 48824}

\author{John Wise$^\dagger$}
\affil{Department of Astrophysical Sciences,  Princeton University, Peyton Hall, Ivy Lane, Princeton, NJ 08544 \\
\hspace*{0.15in}Center for Relativistic Astrophysics, Georgia Institute of Technology, 837 State Street, Atlanta, GA 30332}
\thanks{$^\dagger$Hubble Fellow}

\begin{abstract}

We describe cosmological galaxy formation simulations with the
adaptive mesh refinement code Enzo that incorporate a star formation
prescription regulated by the local abundance of molecular
hydrogen. We show that this \HH-regulated prescription leads to a
suppression of star formation in low mass halos ($M_{\rm h} \lesssim
10^{10} \msun$) at $z>4$, alleviating some of the dwarf galaxy
problems faced by theoretical galaxy formation models. \HH\ regulation
modifies the efficiency of star formation of cold gas directly, rather
than indirectly reducing the cold gas content with ``supernova
feedback''. We determine the local \HH\ abundance in our most refined
grid cells (76 proper parsec in size at $z=4$) by applying the model
of Krumholz, McKee, \& Tumlinson, which is based on idealized 1D
radiative transfer calculations of \HH\ formation-dissociation balance
in $\sim 100$ pc atomic--molecular complexes. Our \HH-regulated
simulations are able to reproduce the empirical (albeit lower $z$)
Kennicutt-Schmidt relation, including the low $\SigmaGas$ cutoff due
to the transition from atomic to molecular phase and the metallicity
dependence thereof, without the use of an explicit density threshold
in our star formation prescription. We compare the evolution of the
luminosity function, stellar mass density, and star formation rate
density from our simulations to recent observational determinations of
the same at $z=4-8$ and find reasonable agreement between the two.

\end{abstract}
 
\section{Introduction}

The $\Lambda$CDM paradigm of cosmological structure formation
\citep{white_core_1978,blumenthal_formation_1984} has been
tremendously successful at explaining the large scale, statistical
features of the distribution of matter in our universe
\citep{springel_large-scale_2006}. At the same time, it is also very
clear that the mapping from dark matter halos to their baryonic
components, to the properties of galaxies embedded within the halos,
is far from straightforward and currently poorly understood. One
prominent example of this lack of understanding is the fact that the
cosmic mass-to-light relation is neither constant nor monotonic, and
instead exhibits a minimum at a galaxy mass of $10^{12} \msun$
\citep[e.g.][]{conroy_connecting_2009}. Evidently some unknown processes are
inhibiting efficient star formation on both larger and smaller mass
scales.


The focus of this paper is on the low mass end, the dwarf
galaxies. There are (at least) two dwarf galaxy problems, which may or
may not have the same explanation. The first of these is the
well-known ``Missing Satellites Problem''
\citep{kauffmann_formation_1993,klypin_where_1999,moore_dark_1999},
which refers to the discrepancy between the relatively small number of
satellite galaxies known to be orbiting the Milky Way and M31 ($\sim
20$ around each) and the vastly larger number of dark matter subhalo
satellites predicted from dark-matter-only cosmological numerical
simulations \citep[$\gtrsim 10^5$ in the latest simulations,][]{diemand_clumps_2008,stadel_quantifying_2009,springel_aquarius_2008}. Photo-heating
from the meta-galactic UV background will prevent gas from collapsing
and forming stars in all but the most massive subhalos
\citep{efstathiou_suppressing_1992,kauffmann_formation_1993,bullock_reionization_2000},
but even so the number of dark matter subhalos that should be able to
host a luminous component, either because they collapsed prior to
reionization or because they reached a sufficiently large mass
thereafter, exceeds the current census of dwarf satellite galaxies by
at least one order of magnitude \citep{madau_dark_2008}. The
interpretation of this discrepancy is further complicated by
interactions with the host galaxy. Ram pressure stripping
\citep{mayer_simultaneous_2006}, as well as resonant
\citep{donghia_resonant_2009} and tidal interactions
\citep{gnedin_tidal_1999} with the host's dark matter halo or stellar
disk could modify the abundance and properties of galactic satellite
galaxies.

The second dwarf galaxy problem occurs in the field, and is
exemplified by the apparent inability of virtually all theoretical
models of galaxy formation to match the abundance of low stellar mass
galaxies at $z>0$. Semi-analytic galaxy formation models (SAMs), for
example, are able to match the observed stellar mass function in the
local universe ($z=0$) by judiciously tuning their AGN,
photo-ionization, and supernova feedback parameters, but these same
models predict an abundance of $M_\star < 10^{10} \msun$ galaxies at
higher redshifts that exceeds the observational constraints by an
order of magnitude
\citep{fontanot_many_2009,marchesini_evolution_2009,cirasuolo_new_2010}.
Hydrodynamic galaxy formation simulations face similar problems
\citep{nagamine_history_2006,cen_where_2006,choi_inconsistency_2011}. A closely related problem is the inability of numerical simulations and SAMs to match the low values of the stellar mass fraction and its strongly decreasing trend with halo mass, as inferred from observations and semi-empirical approaches \citep{guo_how_2010,avila-reese_specific_2011}.

Besides UV photo-heating, stellar feedback, in the form of energy or
momentum injection from supernovae explosions
\citep{dekel_origin_1986,efstathiou_model_2000}, stellar winds
\citep{norman_clumpy_1980,mckee_photoionization-regulated_1989}, or
radiation pressure
\citep{krumholz_dynamics_2009,murray_disruption_2010,hopkins_self-regulated_2011},
is commonly invoked to explain the reduction in star formation
efficiency in low mass halos. In SAMs
\citep{cole_hierarchical_2000,benson_what_2003,somerville_semi-analytic_2008,wang_dependence_2008}
this type of feedback is typically modeled as a reduction of the cold
gas reservoir available for star formation, with an efficiency
proportional to some power law of the galactic disk circular velocity.

In direct cosmological numerical simulations, stellar feedback remains
subgrid physics even for today's state-of-the-art computational
efforts. Cosmological zoom-in simulations of individual galaxies have
reached tens of parsec resolution
\citep{gnedin_modeling_2009,ceverino_role_2009,governato_bulgeless_2010,agertz_formation_2011,faucher-giguere_small_2011,guedes_forming_2011},
but full-box global simulations are at least one order of magnitude
behind
\citep{ocvirk_bimodal_2008,schaye_physics_2010,oppenheimer_feedback_2010,choi_inconsistency_2011,faucher-giguere_baryonic_2011}. Neither
approach is able to resolve the $\lesssim$ parsec scales on which
stellar feedback actually operates in nature. Instead simulators have
turned to feedback prescriptions that are meant to capture the
cumulative effect of supernovae explosions on scales that are
computationally accessible. A wide variety of such subgrid physics
prescriptions have been implemented, ranging from a simple injection
of thermal energy at the location of newly created star particles
\citep{cen_cold_1993}, often with radiative cooling artificially
turned off for some time to prevent the newly added energy from
rapidly radiating away
\citep{thacker_implementing_2000,stinson_star_2006}, to attempts at
keeping track of separate cold and hot phases of the subgrid
interstellar medium
\citep{yepes_hydrodynamical_1997,gnedin_metal_1998,springel_cosmological_2003},
to direct kinetic feedback, in which momentum kicks are applied to
surrounding gas particles
\citep{springel_cosmological_2003,oppenheimer_cosmological_2006,schaye_physics_2010,genel_short-lived_2010},
which are subsequently temporarily decoupled from hydrodynamic forces
\citep[except in][]{schaye_physics_2010} in order to allow them to
escape the star forming region.

Although improvements in the stellar feedback treatment have indeed
enabled progress in galaxy formation simulations, for example the
production of quasi-realistic disk galaxies from cosmological initial
conditions
\citep{governato_bulgeless_2010,agertz_formation_2011,guedes_forming_2011,avila-reese_specific_2011,piontek_modelling_2011,brook_magicc_2012},
the results depend sensitively on the details of the feedback
implementations \citep{sales_feedback_2010}, which themselves are
often based on ad-hoc assumptions. Furthermore, many problems are not
completely solved by stellar feedback as it is currently
implemented. The inability of simulations and SAMs, even those
including supernova feedback prescriptions, to match the observed high
redshift stellar mass functions and star formation rates is one
example \citep{cirasuolo_new_2010,choi_inconsistency_2011}. Another is
the challenge of suppressing the stellar mass content of low mass
halos while simultaneously matching the observed mass-metallicity
relation in Milky Way dwarf satellites \citep{font_new_2011}.

Important physical processes associated with star formation are not
captured by current models, and it is time to revisit the dwarf galaxy
problems in light of new understanding of how SF actually occurs in
the ISM of local galaxies. One promising direction is an improved
treatment of the chemistry and thermodynamics of the interstellar gas
that is actually forming stars. In particular, spatially resolved
observations of local galaxies have revealed that star formation
correlates much more tightly with the density of \textit{molecular}
gas than total gas density
\citep{wong_relationship_2002,kennicutt_star_2007,leroy_star_2008,bigiel_star_2008}. Even
though the primary cooling agents are lines of CO or CII (depending on
the chemical state of the carbon), molecular hydrogen (\HH) is
expected to be good tracer of star formation, even at low
metallicities \citep{krumholz_which_2011}.

This motivates a star formation prescription that differentiates
between the chemical phases of the gas, in which the formation of star
particles is tied to the local abundance of \HH, as opposed to the
total gas density, as is more commonly done in numerical
simulations. Indeed some SAMs have begun to explore this direction
\citep[Krumholz \& Dekel
  2011]{fu_atomic--molecular_2010,lagos_impact_2010}, and several
numerical simulations including \HH\ physics have been published
\citep{robertson_molecular_2008,gnedin_modeling_2009,feldmann_how_2011}.
Until now these simulations focused only on individual galaxies,
either in isolated disks or in cosmological zoom-in simulations. In
this work we investigate for the first time the effect of an
\HH-regulated star formation prescription in full-box cosmological
simulations, with an eye towards the statistical distribution of star
formation efficiency in dwarf galaxies.

Following the non-equilibrium \HH\ chemistry in a realistic and
self-consistent manner, including formation on dust grains and the
radiative transfer of ionizing and dissociating radiation, is
complicated and expensive to implement in numerical galaxy formation
simulations \citep{gnedin_modeling_2009}. Fortunately, analytical 1D
radiative transfer calculations assuming \HH\ formation-dissociation
balance
\citep{krumholz_atomic--molecular_2008,krumholz_atomic--molecular_2009,mckee_atomic--molecular_2010}
have shown that the \HH\ abundance is determined to a good
approximation \citep{krumholz_comparison_2011} by the HI column
density and metallicity of gas on $\sim 100$ pc scales. As these
scales are directly accessible to us, we can bypass much of the
computational difficulty associated with proper \HH\ chemistry by
implementing the \citet{krumholz_atomic--molecular_2009} results in
our cosmological simulations. Even so, our use of cosmological
adaptive mesh refinement prevents our simulations from progressing
much past $z \sim 4$ at an acceptable computational expense. Yet it is
precisely in the early universe, at low but non-zero metallicities,
that the metallicity-dependence of the HI to \HH\ transition will be
most important, and its effect on star formation greatest. Thanks to
extensive multi-wavelength surveys \citep[e.g. the Great Observatories
Origins Deep Survey,][]{giavalisco_great_2004} and deep follow-up
observations with the Hubble and Spitzer space telescopes
\citep[e.g.][]{stark_evolutionary_2009,bouwens_ultraviolet_2011,gonzalez_evolution_2011,labbe_ultradeep_2010},
we are able to make contact with observational constraints on the
cosmic stellar mass and star formation density even at these high
redshifts.

The main driver of this work, then, is to investigate to what degree a
proper accounting of the \HH\ abundance in star forming gas can take
the role that is traditionally assigned to supernova feedback, namely
a reduction of the star formation efficiency in low mass dark matter
halos \citep[see also][hereafter G09;
  GK10]{gnedin_modeling_2009,gnedin_kennicutt-schmidt_2010}. This
paper is organized as follows. In \S~\ref{sec:simulations} we
describe our numerical approach and the details of our \HH-regulated
star formation prescription. We present the results of our work in
\S~\ref{sec:Kennicutt-Schmidt} - \ref{sec:evolution}. We first
show that our prescription reproduces many of the observational
features of the star formation scaling relations
(\S~\ref{sec:Kennicutt-Schmidt}). We then demonstrate that tying
star formation to the \HH\ abundance indeed suppresses star formation
in low mass halos, thereby alleviating the dwarf galaxy problems
(\S~\ref{sec:fstar}). Finally, we present a direct comparison to
recent observational determinations of the high redshift evolution of
the luminosity function, stellar mass density, and star formation rate
density (\S~\ref{sec:evolution}). We summarize and conclude in
\S~\ref{sec:conclusions}.

\section{Simulations}
\label{sec:simulations}

We have conducted cosmological AMR hydrodynamics simulations using
Enzo v2.0\footnote{http://code.google.com/p/enzo/} to follow galaxy
formation in the early ($z \geq 4$) universe. The computational domain
covers a (12.5 Mpc)$^3$ box with a root grid of $256^3$ grid
cells. The dark matter density field is resolved with $256^3$
particles of mass $3.1 \times 10^6 \msun$. The box has mean over-density of zero, and no additional density fluctuation on the scale of the simulation box \citep[``DC mode'',][]{gnedin_implementing_2011} has been applied. Adaptive mesh refinement is
allowed to occur throughout the entire domain for a maximum of 7
levels of refinement, resulting in a maximum spatial resolution of
$\Delta x_7 = 76.3 \times 5/(1+z)$ proper parsec. Mesh refinement is
triggered by a grid cell reaching either a dark matter mass equal to 4
times the mean root grid cell dark matter mass, or a baryonic mass
equal to $8 \times 2^{-0.4 l}$ times the mean root grid cell baryonic
mass, where $l$ is the grid level. The negative exponent in the
baryonic refinement mass threshold implies a super-Lagrangian
refinement criterion that results in more aggressive refinement at
higher resolution. The simulations are initialized at $z_i$=99 with
the \citet{eisenstein_power_1999} transfer function, and cosmological
parameters consistent with the WMAP 7-year results
\citep{komatsu_seven-year_2011}: $\Omega_M = 0.265$, $\Omega_\Lambda =
0.735$, $\Omega_b h^2 = 0.02264$, $h=0.71$, $n=0.963$, and
$\sigma_8=0.801$. The parameters of our suite of simulations are
summarized in Table~\ref{tab:SF_parameters}.

The equations of hydrodynamics are solved using Enzo's implementation
of the Piecewise Parabolic Method \citep[PPM,][]{colella_piecewise_1984}, a higher order accurate
Godunov scheme. We utilize the recently added HLLC Riemann solver
\citep{toro_restoration_1994} with a fallback scheme to the more
diffusive HLL solver for problematic cells, which greatly aids the
simulations' stability. Enzo employs a dual-energy formalism
\citep{bryan_piecewise_1995}, solving for both the internal gas energy
and total energy separately, to ensure accurate pressures and
temperatures in hypersonic flows.

Our simulations include radiative cooling from both primordial and
metal enriched gas, as well as photo-heating from an optically thin,
uniform meta-galactic UV background. The primordial gas cooling rates
are calculated from the ionization states of hydrogen and helium,
which are followed with a 6-species (H, H$^+$, He, He$^+$, He$^{++}$,
and e$^-$) non-equilibrium chemical network
\citep{abel_modeling_1997,anninos_cosmological_1997}, including
collisional and photo-ionization/excitation rates. The metal cooling
is determined from a 5-dimensional table (independent variables:
density, temperature, electron fraction, metallicity, and redshift) of
heating and cooling rates precomputed with the Cloudy code
\citep{ferland_cloudy_1998}, as described in detail in
\citet{smith_metal_2008} and \citet{smith_nature_2011}. For the UV
background we used the updated version of the
\citet{haardt_modelling_2001} UV background model that ships with
version 07.02.01 of Cloudy. This model includes the contributions of
both quasars and galaxies in a redshift dependent manner, and is in
reasonable agreement with a more recent calculation of the cosmic UV
background \citep{faucher-giguere_new_2009}.

As is commonly done in Eulerian hydrodynamic galaxy formation
simulations
\citep{machacek_simulations_2001,robertson_molecular_2008,agertz_disc_2009,ceverino_role_2009},
we apply an artificial pressure support to cells that have reached the
maximum refinement level. This is necessary in order to stabilize
these cells against artificial fragmentation, and is supposed to mimic
the pressure support from turbulent motions below the simulation's
resolution. In Enzo this support is implemented by increasing the
internal gas energy up to some multiple of the value required to make
the cell Jeans stable. We have set this factor equal to 10, meaning
that the Jeans length of the highest resolution cells is artificially
set to $\sqrt{10} \approx 3$ times the cell width.

\begin{deluxetable*}{lcccccc}
\tablecaption{Summary of the simulations.}
\tablewidth{0pt}
\tablehead{
\colhead{Name} & \colhead{$z_{\rm final}$} & \colhead{$\rho_{\rm gas, SF}$} & \colhead{$n_{\rm thresh}$} & \colhead{$J_{\rm LW}/J_{\rm MW}$} & \colhead{$[\Zfloor]$} & \colhead{Comment}
}
\startdata
KT07 & 4.0 & tot & $50 \, {\rm cm}^{-3}$ & | & | & \citet{krumholz_slow_2007} SF law \\
KT07\_low & 6.0 & tot & $5 \, {\rm cm}^{-3}$ & | & | & lower SF threshold \\
KT07\_high & 6.0 & tot & $500 \, {\rm cm}^{-3}$ & | & | & higher SF threshold \\
KMT09 & 4.0 & \HH &  | & | & -3.0 & \citet{krumholz_atomic--molecular_2009}: 2-phase equilibrium  \\
KMT09\_L8 & 6.0 & \HH &  | & | & -3.0 & one additional refinement level (maxlevel=8)  \\
KMT09\_FLW1 & 5.0 & \HH & | & 1 & -3.0 & KMT09 with uniform LW\\
KMT09\_FLW10 & 5.0 & \HH & | & 10 & -3.0 & background of \\
KMT09\_FLW100 & 5.0 & \HH & | & 100 & -3.0 &  increasing \\
KMT09\_FLW1000 & 5.0 & \HH & | & 1000 & -3.0 & intensity \\
KMT09\_ZF4.0 & 6.0 & \HH & | & | & -4.0 & lower $\Zfloor$ \\
KMT09\_ZF2.5 & 6.0 & \HH & | & | & -2.5 & higher $\Zfloor$ \\
KMT09\_ZF2.0 & 6.0 & \HH & | & | & -2.0 & even higher $\Zfloor$ \\
KMT09\_ZFz10 & 6.0 & \HH & | & | & -3.0 & $\Zfloor$ at $z=10$ \\
KMT09\_Sob & 5.0 & \HH &  | & | & -3.0 & Sobolev-like approximation of $\Sigma$ \\
KMT09\_SobL8 & 6.0 & \HH &  | & | & -3.0 & KMT09\_Sob with maxlevel=8
\enddata

\tablecomments{All simulations have the same box size (12.5 Mpc) and
  were initialized at $z=99$ with a WMAP7 cosmology: $\Omega_M=0.265$,
  $\Omega_\Lambda=0.735$, $\Omega_b=0.045$, $h=0.71$,
  $\sigma_8=0.801$, and $n_s=0.963$. The number of dark matter
  particles is $256^3$ ($m_p=3.64 \times 10^6 \msun$), and the root
  grid dimensions are also $256^3$. We allow up to 7 levels of
  adaptive mesh refinement, except in KMT09\_L8 which has one
  additional level. $\rho_{\rm gas, SF}$ indicates whether the SF
  prescription is tied to the total or \HH\ gas density, $n_{\rm
    thresh}$ is the minimum density required for SF to occur (KT07
  runs only), $J_{\rm LW}/J_{\rm MW}$ the intensity of the
  Lyman-Werner background normalized to the Milky Way's value
  (KMT09\_FLW runs only), and $[\Zfloor] \equiv
  \log_{10}(\Zfloor/\Zsun)$ is the initial seed metallicity applied at
  $z=9$ in the KMT09 simulations.}
\label{tab:SF_parameters}
\end{deluxetable*}

\subsection{Star Formation Prescriptions}

The star formation (hereafter SF) prescriptions we have implemented
are all variations on the basic Schmidt law, whereby the local star
formation rate (hereafter SFR) in a grid cell is proportional to its
gas density divided by a SF time scale,
\begin{equation}
\dot{\rho}_{\rm SF} = \epsilon \, \f{ \rhoGas }{ t_\star }.
\end{equation}

In all of our simulations we set the SF time scale equal to the local
free-fall time,
\begin{equation}
t_\star = t_{\rm ff} = \sqrt{\f{3\pi}{32 \,G \, \rhoGas}},
\end{equation} 
and fix the SF efficiency to $\epsilon = 0.01$, as motivated by
\citet{krumholz_slow_2007}, who showed that the local SF efficiency per
free fall time is low ($\epsilon_{\rm SF} \equiv \dot{\SigmaSFR}/(\SigmaGas \, t_{\rm ff}) \approx 0.01$) and
approximately constant over 4 orders of magnitude in density.

In Enzo's standard routines, SF is allowed to occur at every time step
and in every grid cell that is not further refined. Provided the cell
fulfills all conditions for SF, a fraction of the cell's gas mass is
converted into a star particle of mass $m_p = \epsilon \, \rho_{\rm
  gas} \, (\Delta x)^3 (\Delta t/t_\star)$. For highly refined cells
with small time steps $\Delta t$, this commonly result in very large
numbers of low mass star particles, which can dramatically slow down
the simulation's progress. Applying a mass threshold, below which a
star particle is simply not created, is undesirable, since it can lead
to a significant amount of ``unfulfilled'' SF, although this can be
remedied with a stochastic SF criterion \citep[e.g.][and see
  below]{springel_cosmological_2003}.

To overcome these difficulties, we have modified Enzo's routines to
allow SF to occur only \textit{once per root grid time step} and only
in cells at the \textit{highest refinement level} (here $l=7$), but
with a star particle mass proportional to the root grid time step
$\Delta t_0$ \citep[see][]{kravtsov_origin_2003}, i.e.
\begin{equation}
m_p = \epsilon \, \rhoGas \, (\Delta x_7)^3 \, \f{\Delta t_0}{t_\star}.
\end{equation}
Making $m_p$ proportional to the root grid time step ($\Delta t_0 \gg
\Delta t_7$) goes a long way towards overcoming the problem of large
numbers of low mass star particles. Nevertheless we also enforce a
minimum star particle mass of $m_{\rm min} = 10^4 \msun$, since even
$\Delta t_0$ can occasionally become very small. Below this mass we
implement a stochastic SF criterion as follows: if $m_p < m_{\rm
  min}$, we form a particle of mass equal to $m_{\rm min}$ if a
randomly generated number is smaller than $(m_p/m_{\rm min})$.

We consider two distinct classes of SF prescriptions:
\begin{enumerate}[i)]
\item \textit{Standard SFR (KT07)}: The SFR is proportional to the
  \textit{total} gas density divided by the free-fall time, resulting
  in a SFR proportional to $\rho_{\rm gas}^{3/2}$. We apply a density
  threshold below which SF is not allowed to occur, and vary this
  threshold between values of 5, 50, and 500 cm$^{-3}$.
\item \textit{\HH-regulated SFR (KMT09)}: The SFR is proportional to
  the \textit{molecular hydrogen} density divided by the free-fall
  time determined from the total gas density, resulting in a SFR
  proportional to $\fHH \, \rho_{\rm gas}^{3/2}$. The \HH\
  fraction, $\fHH = \rho_{{\rm H}_2}/\rho_{\rm gas}$, is
  determined following \citet{krumholz_atomic--molecular_2009} (more
  details in section~\ref{sec:H2}), and we consider both the two-phase
  equilibrium model and a range of models with different Lyman-Werner
  \HH-dissociating background intensities. No density threshold is
  applied.
\end{enumerate}

The resolution of our simulations is not sufficient to resolve the
formation sites of the first generation of stars, the so-called
population III. In order to capture the metal enrichment resulting
from the supernova explosions of this primordial stellar population,
we instantaneously introduce a metallicity floor of $[\Zfloor] \equiv
\log_{10}(\Zfloor/\Zsun) = -3.0$ at $z=9$, as motivated by recent high
resolution numerical simulations of the transition from Pop.III to
Pop.II SF \citep{wise_birth_2012}. This ensures the presence of a
minimum amount of metals, which seed subsequent star formation and
further metal enrichment. We discuss the sensitivity of our results to
the time and amplitude of this metallicity floor in
\S~\ref{sec:fstar_Zfloor}.

\subsection{Molecular Chemistry}
\label{sec:H2}

To obtain the molecular hydrogen mass fraction $\fHH \equiv \rho_{{\rm
    H}_2} / \rho_{\rm gas}$ in a given grid cell, we follow the
analytical model developed in \citet{krumholz_atomic--molecular_2008},
\citet{krumholz_atomic--molecular_2009}, and
\citet{mckee_atomic--molecular_2010}. This model is based on a
radiative transfer calculation of an idealized spherical giant
atomic--molecular complex, subject to a uniform and isotropic
Lyman-Werner (LW) radiation field. The \HH\ abundance is calculated
assuming formation-dissociation balance. The solution of this problem
can conveniently be expressed in three lines:
\begin{equation}
\fHH \simeq 1 - \f{3}{4} \f{s}{1 + 0.25 s}, \label{eq:KMT09model_begin}
\end{equation}
\begin{equation}
s = \f{\ln(1 + 0.6\chi + 0.01 \chi^2)}{0.6\,\tau_c},
\end{equation}
\begin{equation}
\chi = 71 \left( \f{\sigma_{d,-21}}{\mathcal{R}_{-16.5}} \right) \f{G'_0}{n_{\rm H,0}},
\end{equation}
where $\tau_c$ is the dust optical depth of the cloud,
$\sigma_{d,-21}$ is the dust cross-section per H nucleus to 1000
\AA\ radiation, normalized to a value of $10^{-21}$ cm$^{-2}$,
$\mathcal{R}_{-16.5}$ is the rate coefficient for \HH\ formation on
dust grains, normalized to the Milky Way value of $10^{-16.5}$ cm$^3$
s$^{-1}$ \citep{wolfire_chemical_2008}, $G'_0$ is the ambient UV
radiation field intensity, normalized to the
\citet{draine_photoelectric_1978} value for the Milky Way, and $n_{\rm
  H,0}$ is the volume density of H nuclei in units of cm$^{-3}$. Since
both $\sigma_d$ and $\mathcal{R}$ are linearly proportional to the
dust abundance, their ratio is independent of the gas metallicity.

\citet{krumholz_atomic--molecular_2009} showed that a further
simplification to the model can be made if the ISM is assumed to be in
two-phase equilibrium between a cold neutral medium (CNM) and and a
warm neutral medium (WNM) \citep{wolfire_neutral_2003}. The assumption
of pressure balance between these two ISM components forces the
minimum CNM density to be linearly proportional to the intensity of
the LW radiation field, with only a weak dependence on metallicity:
\begin{equation}
n_{\rm min} \approx \f{31}{1 + 3.1 \, ({\rm Z}/\Zsn)^{0.365}} \, G'_0,
\end{equation}
where $\Zsn$ is the gas phase metallicity in the solar neighborhood,
and we set $\Zsn = \Zsun$ \citep{rodriguez_oxygen_2011} and $\Zsun =
0.0204$. Allowing for the typical CNM density to be somewhat higher
than this minimum value, $n = \phi_{\rm CNM} \, n_{\rm min}$, we get
\begin{equation}
\chi = 2.3 \left( \f{\sigma_{d,-21}}{\mathcal{R}_{-16.5}} \right) \f{1 + 3.1 \, ({\rm Z}/\Zsn)^{0.365}}{\phi_{\rm CNM}},
\end{equation}
which renders $\fHH$ completely independent of the LW
intensity. As in \citet{krumholz_comparison_2011} we set $\phi_{\rm
  CNM}=3$ and $(\sigma_{d,-21}/\mathcal{R}_{-16.5})=1$.

We have conducted one simulation with this two-phase equilibrium
assumption (KMT09), and a range of simulations without it, for which
we instead specify a spatially uniform LW background intensity equal
to 1, 10, 100, and 1000 times the present day Milky Way value of $7.5
\times 10^{-4}$ LW photons cm$^{-3}$ \citep{draine_photoelectric_1978}
(KMT09\_FLW1, ..., KMT09\_FLW1000).  We directly apply these
prescriptions to the highest resolution grid cells in our simulations,
whose size ($\Delta x_7 = 54.5 \times 7/(1+z)$ proper parsec) is
comparable to the physical dimensions of giant atomic--molecular
complexes \citep{blitz_giant_1993}. The dust optical depth is given by
\begin{eqnarray}
\tau_c & = & \Sigma/\mu_H \, \sigma_d \nonumber \\ 
       & \simeq & 0.067 \, \left( \f{\rm Z}{\Zsn} \right) \, \left( \f{\rho \; \Delta x_7}{1 \msun \, {\rm pc}^{-2}} \right), \label{eq:KMT09model_end}
\end{eqnarray}
where $\Sigma = \rho \, \Delta x_7$ is the cell's column density,
$\mu_{\rm H} = 2.3 \times 10^{-24}$ g is the mean mass per H nucleus,
and we have set the dust cross section per H nucleus to be $\sigma_d =
10^{-21} \, ({\rm Z}/\Zsn$) cm$^2$.

The use of $\Delta x_7$ in the calculation of $\Sigma$ introduces an
undesirable explicit resolution dependence in our algorithm, the
effect of which we investigate in \S~\ref{sec:KS_resolution} and
\ref{sec:fstar_resolution}. A Sobolev-like approximation, $\SigmaSob \equiv
\rho \times (\rho / \nabla \rho)$
\citep[e.g. G09;][]{krumholz_comparison_2011}, would remove the
explicit dependence of the algorithm on the width of the finest grid
cells. As discussed in detail in the Appendix, we have conducted
additional simulations using $\SigmaSob$ in the KMT09 prescription,
and these resulted in slightly lower column densities and reduced SF
rates. At densities relevant for star formation ($n > 5$ cm$^{-3}$),
the differences between the two approximations of $\Sigma$ are small,
less than 20\% in the median with a scatter of 0.26 dex, and well
within the range of uncertainty of the parameters of the KMT09
model. The results presented in this paper were obtained with the
simpler and computationally less expensive direct cell-based
approximation using $\SigmaCell \equiv \rho \Delta x_7$.

Regardless of how the surface density of atomic--molecular complexes
is calculated, we should expect some degree of residual resolution
dependence in our simulations, since we don't include the necessary
stellar feedback physics (see below) that regulates the structure of
molecular clouds and provides pressure support against further
collapse. The artificial pressure support mentioned above, which we
apply in order to avoid spurious fragmentation, is resolution
dependent, and hence higher numerical resolution (through additional
refinement levels) will always result in further collapse and higher
densities in our simulations.

The KMT09 model has recently been tested against numerical simulations
that self-consistently follow the formation and destruction of
\HH\ with a non-equilibrium chemical network including time-dependent
and spatially-inhomogeneous 3D radiative transfer of UV and ionizing
radiation \citep{krumholz_comparison_2011}. The analytical model
agrees extremely well with the numerical results whenever the
metallicity is around 1 per cent solar or greater. This agreement
holds for both ``fixed ISM'' simulations in which the metallicity and
radiation field are kept constant and in cosmological simulations in
which the metallicity and radiation field are computed
self-consistently. At metallicities below 1 per cent solar, the
analytical model \textit{overestimates} $\fHH$. Note that for the runs
without the two-phase equilibrium assumption, we follow
\citet{krumholz_comparison_2011} and apply a ``clumping factor'' of 30
to the \HH\ formation rate
(i.e. $(\sigma_{d,-21}/\mathcal{R}_{-16.5})=1/30$) to account for
unresolved density inhomogeneities below our simulations' resolution
limit. We do not use a clumping factor for the two-phase equilibrium
model, because the two-phase model is in effect a direct estimate of
the proper clumping factor. This physically-motivated clumping model
obviates the need for an ad-hoc correction.

\subsection{Feedback} \label{sec:feedback}

Although we explore in this work to what degree an improved treatment
of the ISM's chemical state can replace the need for stellar feedback
in regulating star formation, some form of feedback is necessary even
in our simulations, if only to enrich the gas with metals that promote
the formation of molecular hydrogen. For this purpose we employ a very
simple feedback mechanism that is meant to simultaneously account for
the mass, metals, and thermal energy deposited by winds from massive
stars and core-collapse supernovae. The feedback is applied
instantaneously at the time of formation of the star particle, is
deposited into the $l=7$ grid cell containing the particle, and
consists of the following three components: (i) a fraction $\epsilon_m
= 0.25$ of the star particle's mass is returned as gas, $\Delta m_{\rm
  tot} = \epsilon_m \, m_p$; (ii) the gas is enriched with a metal
yield of $Y=0.02$, $\Delta m_Z = \Delta m_{\rm tot} ( Y (1-Z_p) + Z_p
)$, where $Z_p$ is the metallicity of the star particle and the two
terms correspond to newly and previously enriched ejected material,
respectively; (iii) the thermal energy of the gas cell is increased by
a fraction $\epsilon_{\rm SN} = 10^{-5}$ of the rest-mass energy of
the newly formed star particle, $\Delta E = \epsilon_{\rm SN} \, m_p
c^2$.

This feedback implementation is commonly applied in cosmological
hydrodynamics simulations, but in fact it is known to be
insufficiently strong. The problem with this instantaneous and
localized feedback is that the thermal energy is applied to grid cells
with very high gas density, in which the cooling time is very
short. The injected energy is thus almost immediately lost to
radiative cooling, and the feedback ``fizzles out''
\citep{katz_dissipational_1992}. We acknowledge the shortcoming of our
current feedback implementation, but since we are focusing here on the
global effects of different star formation prescriptions, we defer
attempts at improving the treatment of feedback to future work.

Finally, we caution that nothing in our implementation of molecular
chemistry obviates the need to rely on a subgrid model of star
formation. Our model should be viewed in the same light as other
subgrid models in wide use, e.g. the two-phase model of
\citet{springel_cosmological_2003} or the blast wave model of
\citet{stinson_star_2006}. The main advantage of our new model is that
it incorporates an explicit metallicity-dependence, which both
observations and theory appear to demand, as we discuss in more detail
below.

\subsection{Halo population}

\begin{figure}[htp]
\includegraphics[width=\columnwidth]{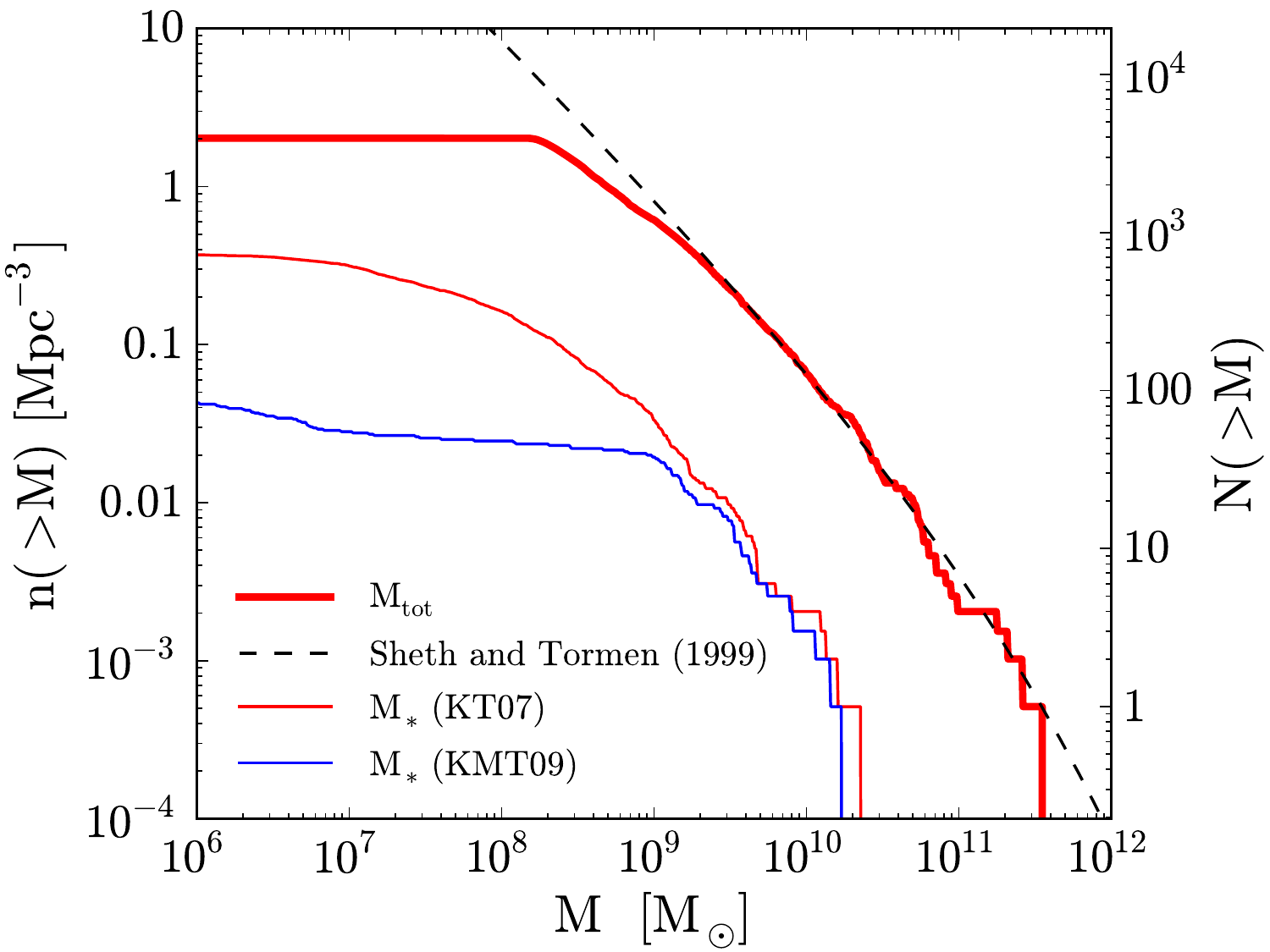}
\caption{The cumulative mass function of halos in our simulations at $z=4$. The thick line is for total mass (KT07 and KMT09 curves are almost indistinguishable, so we only show KT07), and the thin lines are for the stellar mass. The dashed line shows the \citet{sheth_large-scale_1999} mass function fit. The right ordinate gives the total number of objects found in our (12.5 Mpc)$^3$ simulation volume.}
\label{fig:massfunction}
\vspace*{0.1in}
\end{figure}

We used the HOP halo finder \citep{eisenstein_hop:_1998} to identify gravitationally bound dark matter halos in our simulations. For every halo we determined the halo center (defined as the location of the highest dark matter density), the virial radius and corresponding mass (defined as the radius enclosing $\Delta_{\rm vir} \approx $ times the background density \citep{bryan_statistical_1998}), and the amount of gas (total, HI, HII, \HH, HeI, HeII, and metals) contained within the halo.

Fig.~\ref{fig:massfunction} shows the cumulative mass function of the simulations at $z=4$, compared to the \citet{sheth_large-scale_1999} fit. The agreement is remarkably good down to $M \approx 2 \times 10^9 \msun$, corresponding to halos with $\sim 500$ dark matter particles. At even lower mass numerical resolution effects lead to an artifical suppression in the number of halos. A finer root grid resolution and a more aggressive dark matter based refinement criterion would extend the mass function by another order of magnitude in mass \citep{oshea_comparing_2005}. For our purposes, however, a resolution limit of $\sim 2 \times 10^9 \msun$ is sufficient, since halos with $M < 10^9 \msun$ lie below the UV suppression scale \cite{}. Furthermore, as we show in Sec.~\ref{sec:fstar}, \HH\ regulated SF leads to a suppression of stellar mass in halos with $M \lesssim 10^{10} \msun$. As a brief preview of this effect, we overplot in Fig.~\ref{fig:massfunction} the cumulative \textit{stellar} mass functions: the KMT09 simulation has far fewer low stellar mass ($\Mstar < 10^9 \msun$) galaxies, which is a result of the suppression of SF in halos with $M \lesssim 10^{10} \msun$.

\section{The Kennicutt-Schmidt Law}
\label{sec:Kennicutt-Schmidt}

\begin{figure*}[tp]
\includegraphics[width=\textwidth]{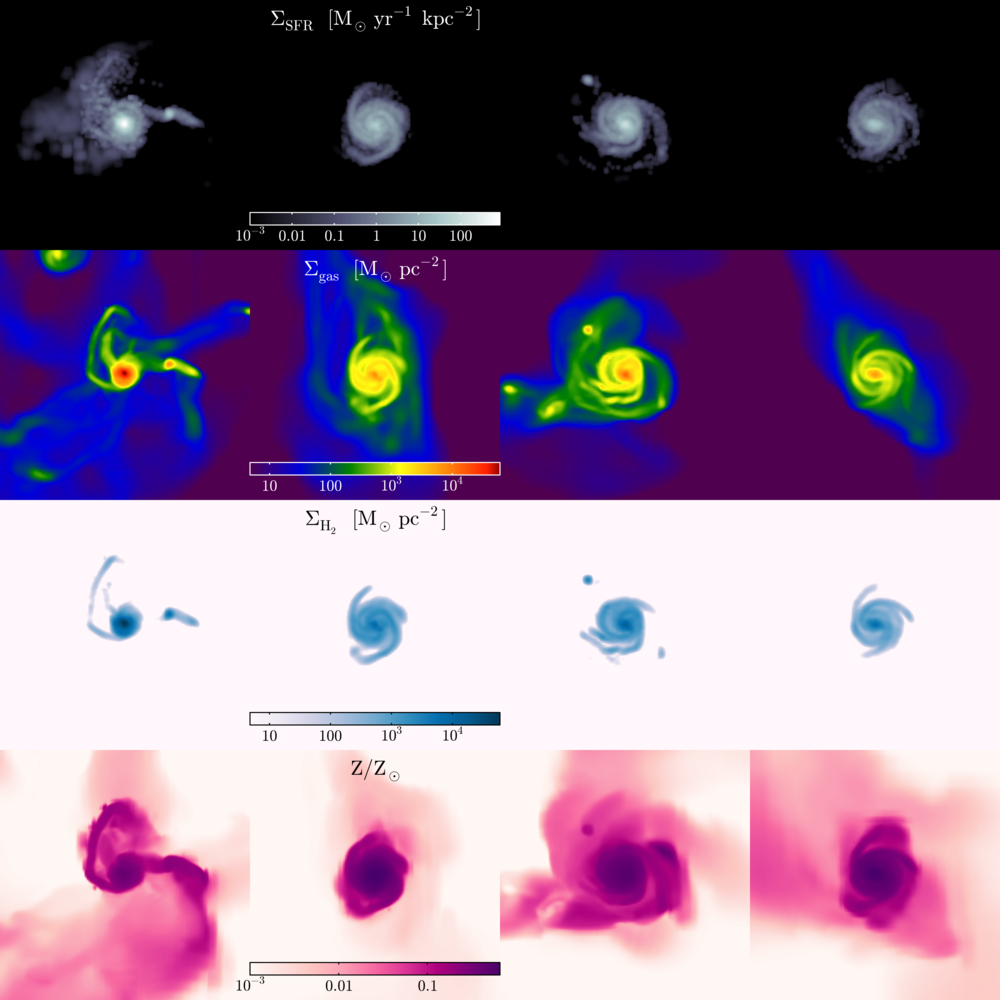}
\caption{Surface density of SFR, total gas, \HH, and metallicity for four representative massive halos ($\Mhalo = 2.5 \times 10^{11}, \, 2.0 \times 10^{11}, \, 1.6 \times 10^{11}, \, 7.4 \times 10^{10} \msun$) in the KMT09 simulation at $z=4$. The projections are calculated as line integrations perpendicular to the disk plane, for a $1000 \times 1000$ pixel grid covering a 10 $\times$ 10 kpc region centered on each galaxy.}
\label{fig:multi_panel}
\vspace*{0.1in}
\end{figure*}

\begin{figure*}[tp]
\includegraphics[width=\textwidth]{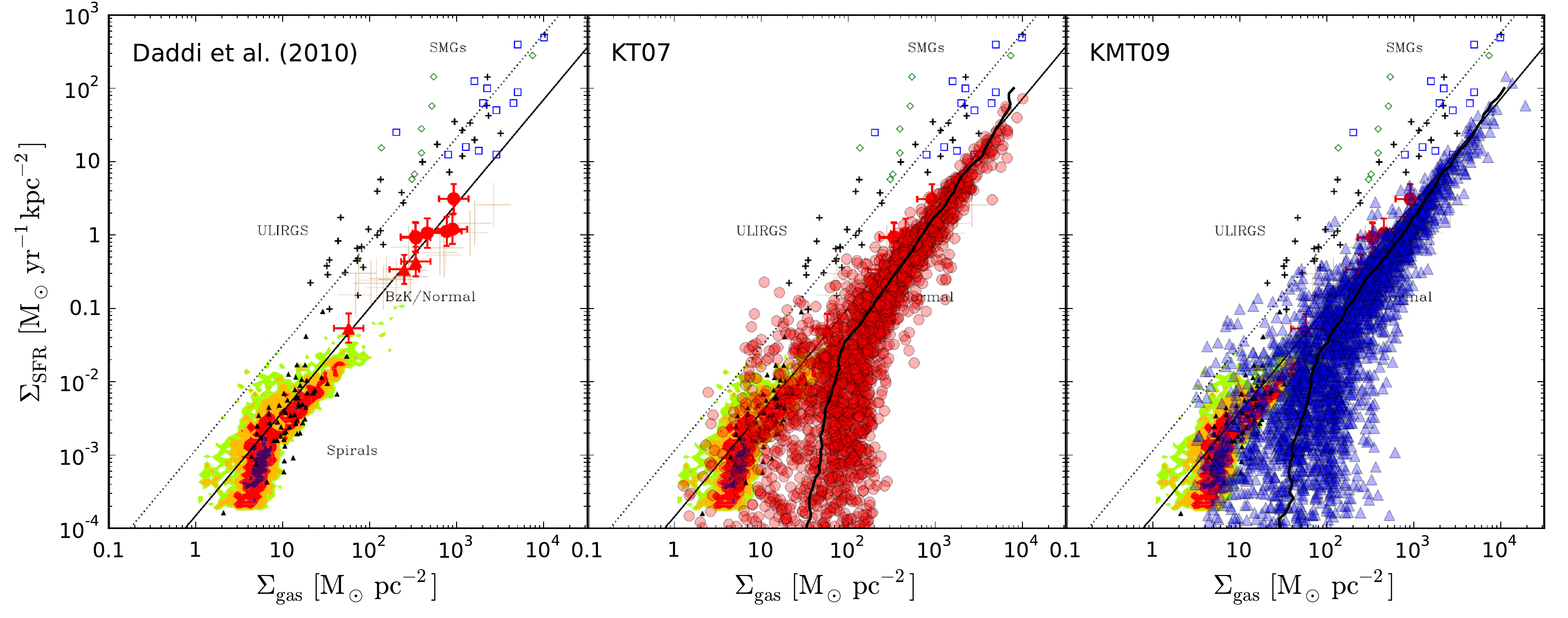}
\caption{Comparison of the observational Kennicutt-Schmidt relation
  from \citet{daddi_different_2010} (left panel) to a standard SF
  simulation with SF density threshold of $n_{\rm thresh} = 50$
  cm$^{-3}$ (middle) and an \HH-regulated SF simulation without any
  density threshold (right) at $z=4$. In the simulations the surface
  densities have been determined from line integrations perpendicular
  to the galaxies' disk plane, and have been smoothed to a resolution
  of 200 pc (see text for more detail).}
\label{fig:Kennicutt}
\vspace*{0.1in}
\end{figure*}
 
\begin{figure}[tp]
\includegraphics[width=\columnwidth]{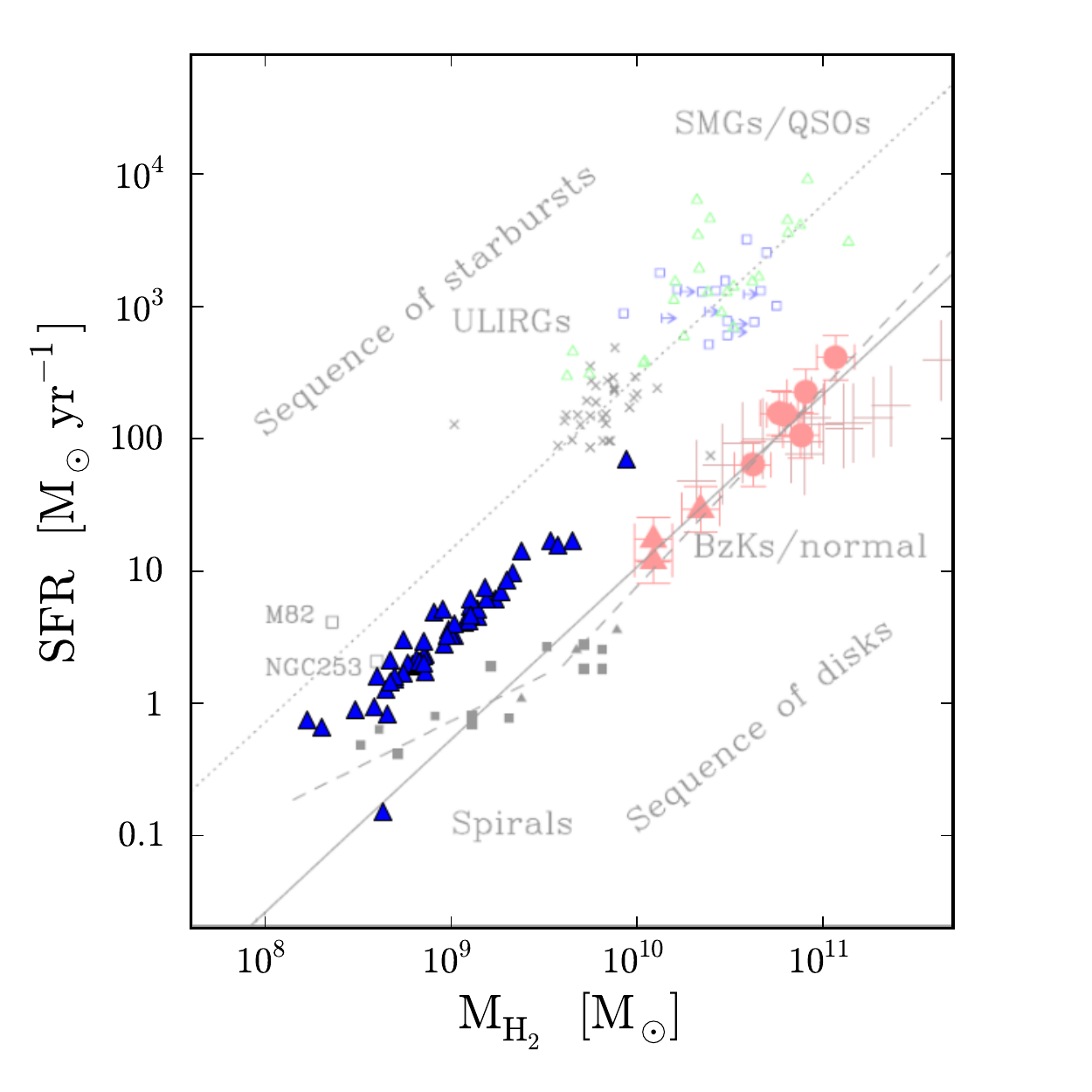}
\caption{Comparison of the relation between total SFR and total \HH\ mass for galaxies in the KMT09 simulation at $z=4$ (blue triangles) with the observational results from \citet{daddi_different_2010} (their Fig.1) for lower redshift galaxies. The simulation and observations have relations with similar slopes, and the simulated galaxies lies in between the ``sequence of starbursts'' and ``sequence of disks''. Higher SFR for a given \HH\ mass may be expected for lower metallicity systems, which require higher densities to allow the transition to \HH\ to occur. }
\label{fig:SFR_vs_MH2}
\vspace*{0.1in}
\end{figure}

A common test for a new numerical SF implementation is to compare the
simulations to the observational Kennicutt-Schmidt (hereafter KS)
relation, the empirical power law between the SFR surface density and
total gas surface density: $\SigmaSFR \propto \SigmaGas^n$, with $n \approx 1.4$
\citep{kennicutt_global_1998,daddi_different_2010,genzel_study_2010}. 

In order to facilitate comparisons to observational data, we determine column densities by integrating through the density fields along the direction perpendicular to the stellar disk of the simulated galaxies. For $\SigmaSFR$ we integrate a SFR density defined as
\begin{equation}
\rho_{\rm SFR} = \sum_{{\rm age}\,<\,\tau_\star} \f{m_\star}{\tau_\star \; (\Delta x)^3}, \label{eq:rhoSFR}
\end{equation}
where $\Delta x$ is the cell width,  $m_\star$ is the star particle mass, and $\tau_\star = 10$ Myr is the SF averaging time scale, which roughly corresponds to observational SFR estimated from nebular emission lines (H$\alpha$, O~III) or FIR continuum, but is a factor of 5 -10 shorter than estimates based on FUV measurements \citep{kennicutt_global_1998,feldmann_how_2011}. The sum in Equation~\ref{eq:rhoSFR} is over all star particles with age less than $\tau_\star$.

We calculate $\SigmaGas$, $\SigmaHH$, $\Sigma_{\rm HI}$, and $\SigmaSFR$ on a $1000 \times 1000$ pixel grid covering a 10 $\times$ 10 kpc region centered on each galaxy in a sample of 35 of the most massive galaxies in our simulations, chosen to cover a wide range in mean metallicity (see \S~\ref{sec:Kennicutt_Z_dependence}). We only considered galaxies in which the adaptive mesh refinement reached the maximum level ($l=7$). Visualizations of $\SigmaSFR$, $\SigmaGas$, $\SigmaHH$, and a density-weighted projection of metallicity  are shown in Figure~\ref{fig:multi_panel} for four representative massive galaxies in the KMT09 simulation at $z=4$.

The observational KS relation, as reported for example
by \citet{daddi_different_2010} and \citet{genzel_study_2010}, has
been established from a wide variety of data out to $z \gtrsim 3$,
including spatially resolved local $z=0$ spiral galaxies,
infrared-selected starbursting galaxies and (U)LIRGs, BzK-selected
galaxies at $z \approx 1.5$, ``normal'' star-forming galaxies at
$z=1-2.3$, and starbursting sub-millimeter galaxies (SMGs) at $z
\approx 1 - 3$ (see \citet{kennicutt_global_1998},
\citet{daddi_different_2010}, and \citet{genzel_study_2010} for
references). A reproduction of the observational relation from
\citet{daddi_different_2010} (their Fig.2) is shown in the left panel
of Figure~\ref{fig:Kennicutt}.

Our simulated galaxies are dwarf galaxies with stellar masses less
than $10^{10} \msun$ at high redshift $z \geq 4$, and hence not
directly comparable to any of these observational galaxy
samples. Nevertheless, since SF is a local process, it makes sense to
directly compare our simulated KS relation to the observations, as a
test of how well our SF prescription is performing. The center and
right panels of Figure~\ref{fig:Kennicutt} show this comparison for
the KT07 and KMT09 simulations at $z=4$. Since the 10 pc intrinsic resolution of our surface density maps is finer than in most spatially resolved studies to date \citep[e.g.][]{bigiel_star_2008,bolatto_state_2011}, we downgrade the spatial resolution to 200 pc by spatially averaging with a 20-cell boxcar average. We explore the resolution dependence of our KS relations in \S~\ref{sec:Kennicutt_res_dependence}. Only pixels with non-zero $\SigmaSFR$ are plotted. The solid line represents a sliding average of $\log_{10} \SigmaGas$ in $\log_{10} \SigmaSFR$ bins of width 0.5.

Both the amplitude and the slope of the relation are qualitatively in
very good agreement with the observations. Such agreement has
previously been reported
\citep{kravtsov_origin_2003,gnedin_kennicutt-schmidt_2010,gnedin_environmental_2011,feldmann_how_2011},
and we show it here merely to demonstrate that our new SF algorithm is
valid and that the use of our \HH-regulated SF prescription does not
destroy this agreement. We note that \citet{kravtsov_origin_2003}
showed that even a \textit{linear} local SF law, with a constant SF
time scale $t_\star$ and a density threshold, can, when spatially
averaged on $\lesssim$ kpc scales, result in a super-linear surface
density relation in agreement with the empirical KS law. Matching the
observed KS law in cosmological simulations is thus not a good way to
distinguish between different SF implementations.

Another cross check with observational data is provided by the demonstration in Fig.~\ref{fig:SFR_vs_MH2} that our simulated galaxies exhibit a similar relation between their SFR and total \HH\ content as the galaxies in the study by \citet{daddi_different_2010}. Our simulated relation has the same slope, and lies in amplitude between what they refer to as the ``sequence of starbursts'' and the ``sequence of disks''. Note that a somewhat higher SFR at a given \HH\ mass may be expected for lower metallicity systems, for which the transition to \HH\ occurs at higher column densities.

\subsection{Low $\SigmaGas$ SF Threshold}
\label{sec:Kennicutt_thresh}

\begin{figure}[tp]
\includegraphics[width=\columnwidth]{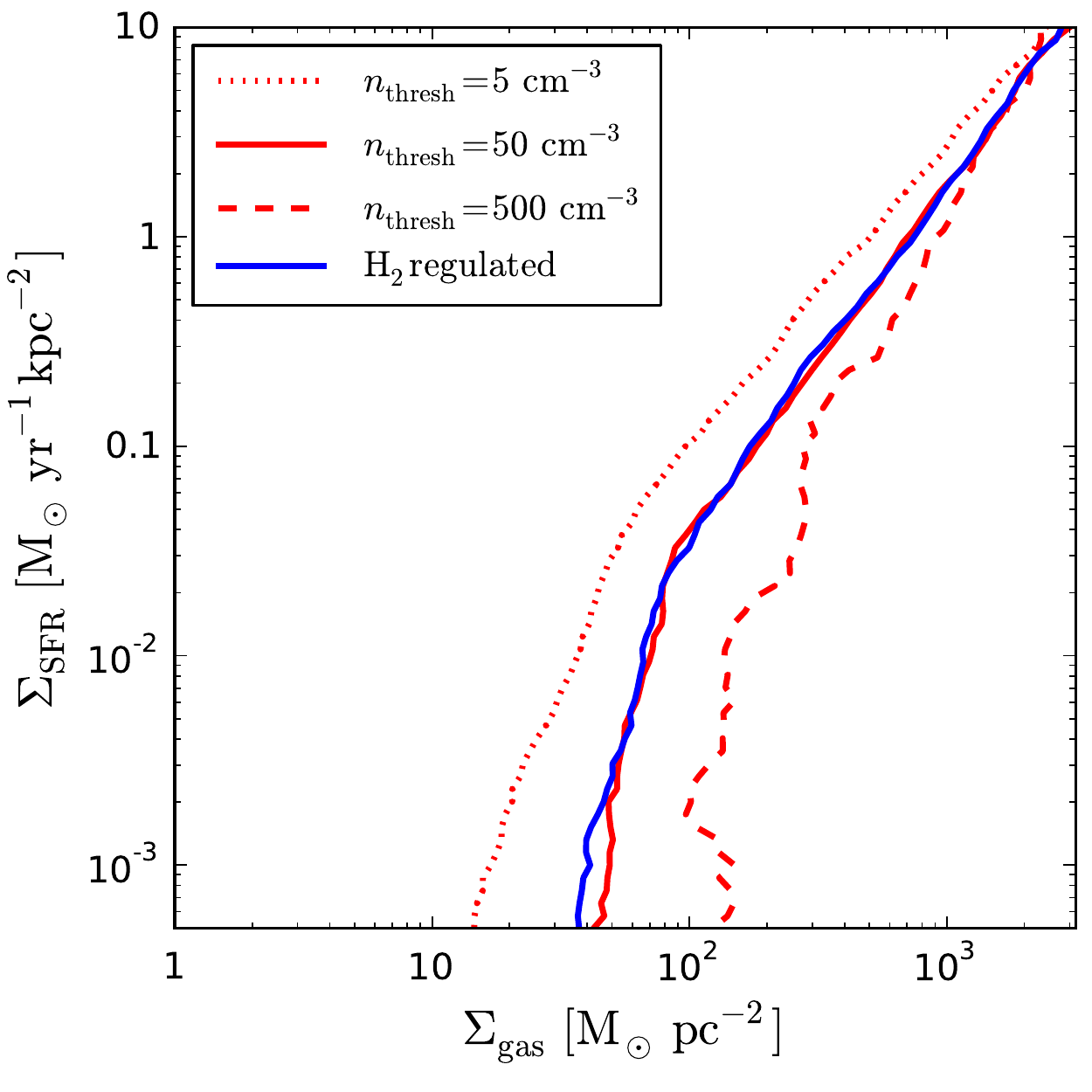}
\caption{Simulated mean Kennicutt-Schmidt relations for standard SF
  with three different density thresholds, $n_{\rm thresh} = 5, \; 50,
  \; 500$ cm$^{-3}$, and the \HH-regulated SF. The KT07 and KMT09
  lines are from the $z=4$ output, the KT07\_low and KT07\_high ones
  from $z=6$.}
\label{fig:Kennicutt_nthresh}
\vspace*{0.1in}
\end{figure}

Both KT07 and KMT09 simulations exhibit a drop off in $\SigmaSFR$ at
$\SigmaGas \approx 50 \msun$ pc$^{-2}$. This feature is believed to
correspond to saturation in the atomic hydrogen fraction, with gas
becoming primarily molecular at higher surface densities
\citep{bigiel_star_2008}. Since $\SigmaSFR$ appears to be almost
independent of $\Sigma_{\rm HI}$ and instead correlates primarily with
$\SigmaHH$, this HI saturation point is reflected in a kink towards
lower SFR in the KS relation.

In the KT07 simulation, which doesn't account for the atomic to
molecular hydrogen transition, this cutoff is reproduced by the
density threshold imposed on the SF. In
Figure~\ref{fig:Kennicutt_nthresh} we show how the simulated KS
relation depends on the value of this threshold. As expected the
relation extends to lower $\SigmaGas$ for the lower threshold $n_{\rm
  thresh}=5$ cm$^{-3}$ case and steepens at a larger value of
$\SigmaGas \approx 100 \msun$ pc$^{-2}$ when the threshold is higher,
$n_{\rm thresh}=500$ cm$^{-3}$ \citep[see also][]{colin_low-mass_2010}. Once the SF is regulated by \HH,
however, this turnover arises naturally: although the curve is noisier
due to the smaller statistics, the relation appears to naturally
steepen at $\SigmaGas \approx 50 \msun$ pc$^{-2}$. This confirms the
earlier results found by GK10 for a set of cosmological zoom-in galaxy
formation simulations including non-equilibrium \HH\ formation and
radiative transfer.

\begin{figure}[tp]
\includegraphics[width=\columnwidth]{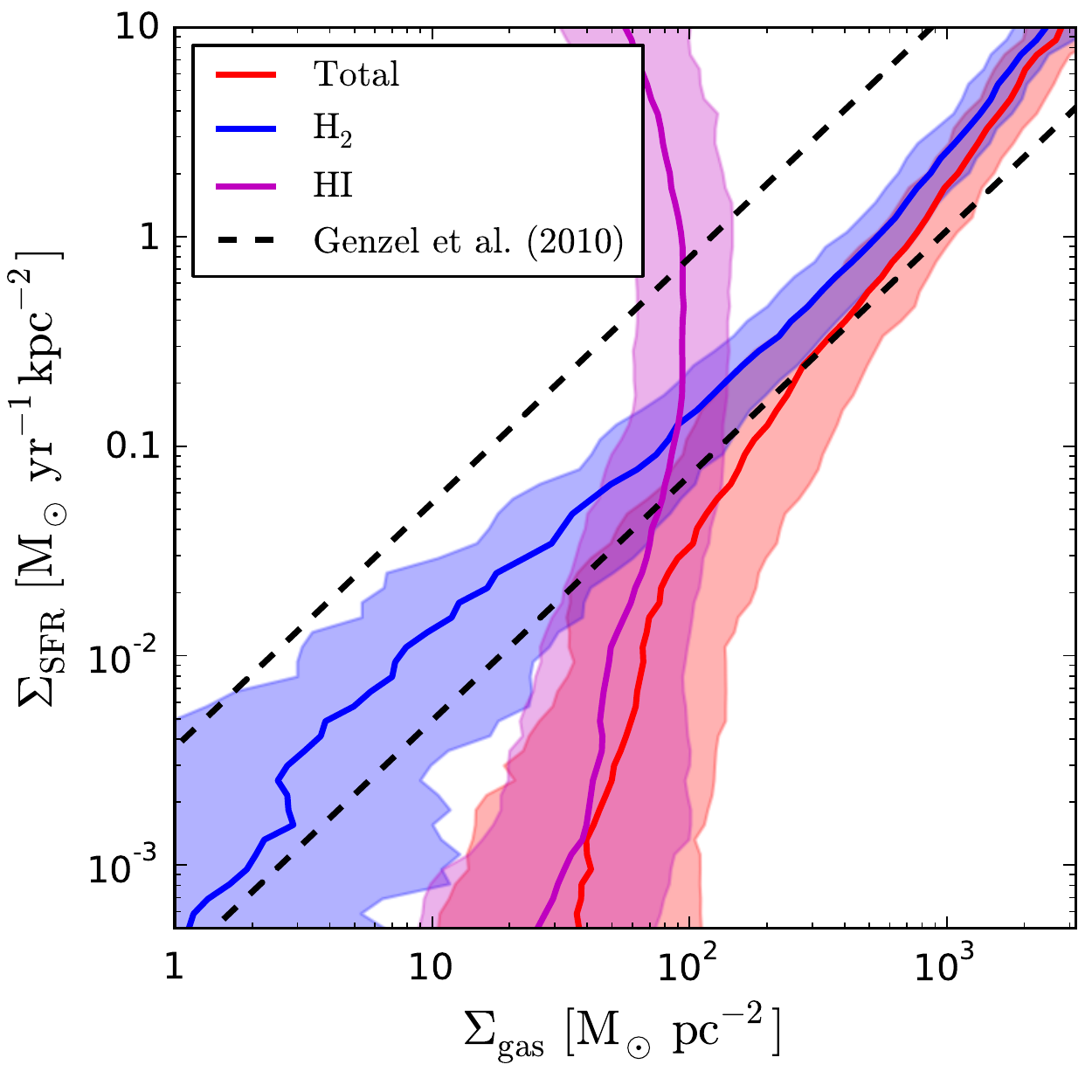}
\caption{Kennicutt-Schmidt relation for total (red), \HH\ (blue), and
  HI (magenta) surface density from the KMT09 simulation at $z=4$. The
  solid lines indicate the mean relations from the simulations, the
  shaded regions cover the central 68\% scatter (16$^{\rm th} -
  84^{\rm th}$ percentile), and the two dashed lines are the best-fit
  $\SigmaHH\!-\SigmaSFR$ relations reported by
  \citet{genzel_study_2010} for their $z = 0 - 3.5$ samples of ``normal'' SF
  galaxies (lower line) and luminous mergers (LIRGs/ULIRGs and SMGs)
  (upper line).}
\label{fig:Kennicutt_H2}
\vspace*{0.1in}
\end{figure}

Note that in the KMT09 KS relation (right panel of
Fig.~\ref{fig:Kennicutt}) there are a few points with very low
$\SigmaGas$ yet non-zero $\SigmaSFR$, and a roughly corresponding
number of points with high $\SigmaGas$ yet values of $\SigmaSFR$
significantly below the KS relation. These points arise from star
particles having wandered out of the high density cells in which they
were born into a neighboring pixel with much lower
$\SigmaGas$. The degree to which this wandering causes a smearing in
the KS relation depends on the spatial and temporal averaging scales
employed in calculating $\SigmaGas$ and $\SigmaSFR$, and on the stars'
velocity dispersion.

\begin{figure}[tp]
\includegraphics[width=\columnwidth]{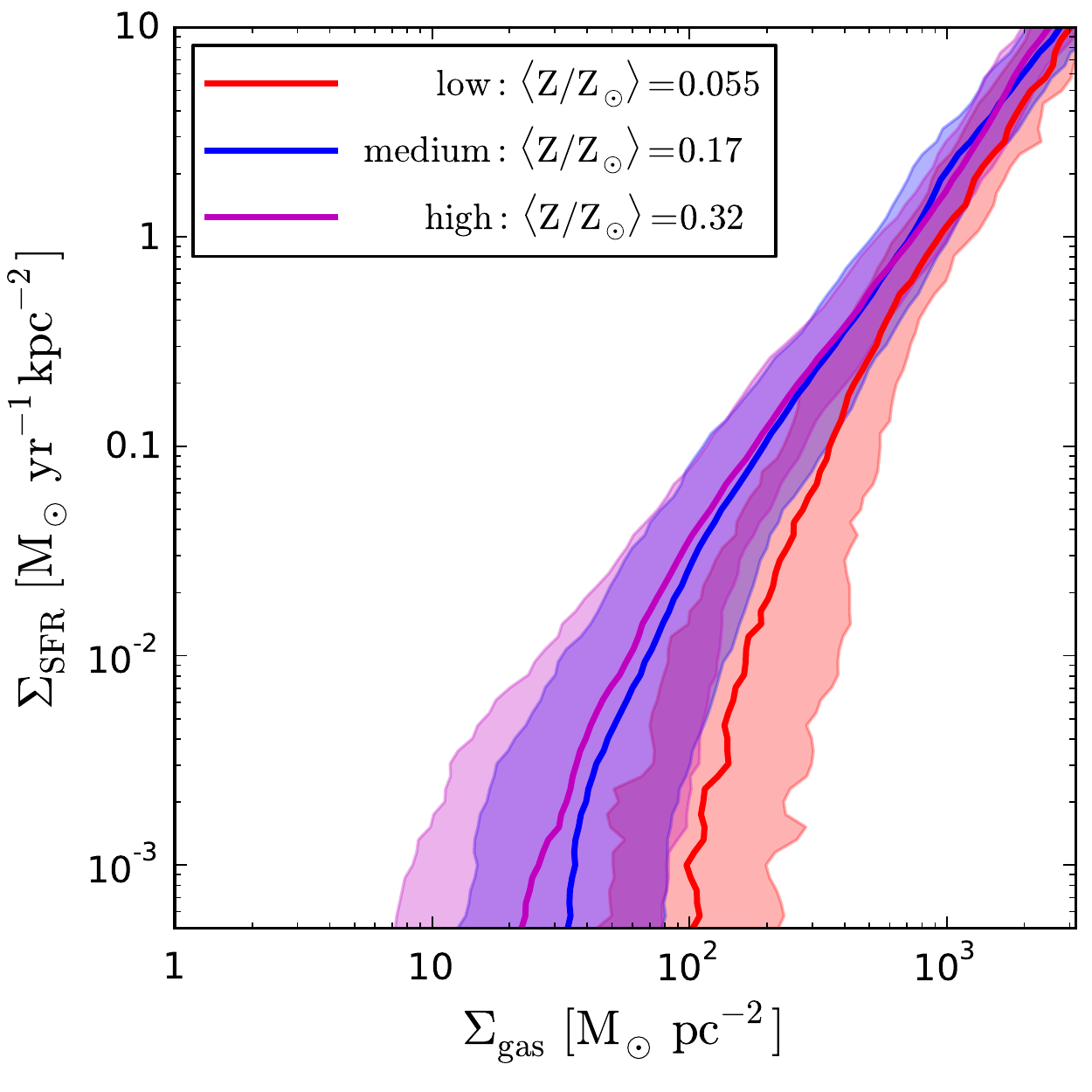}
\caption{The total gas KS relation for simulated galaxy subsamples
  split by their mean gas-phase metallicity, from the KMT09 simulation at
  $z=4$. Solid lines indicate the mean relations from the simulations,
  and the shaded regions cover the central 68\% scatter (16$^{\rm th}
  - 84^{\rm th}$ percentile). The lower the metallicity, the higher
  the $\SigmaGas$ that is needed for the transition to fully molecular
  gas, which corresponds to the turn over in the KS relation.}
\label{fig:Kennicutt_Z}
\vspace*{0.1in}
\end{figure}

\begin{figure}[htp]
\includegraphics[width=\columnwidth]{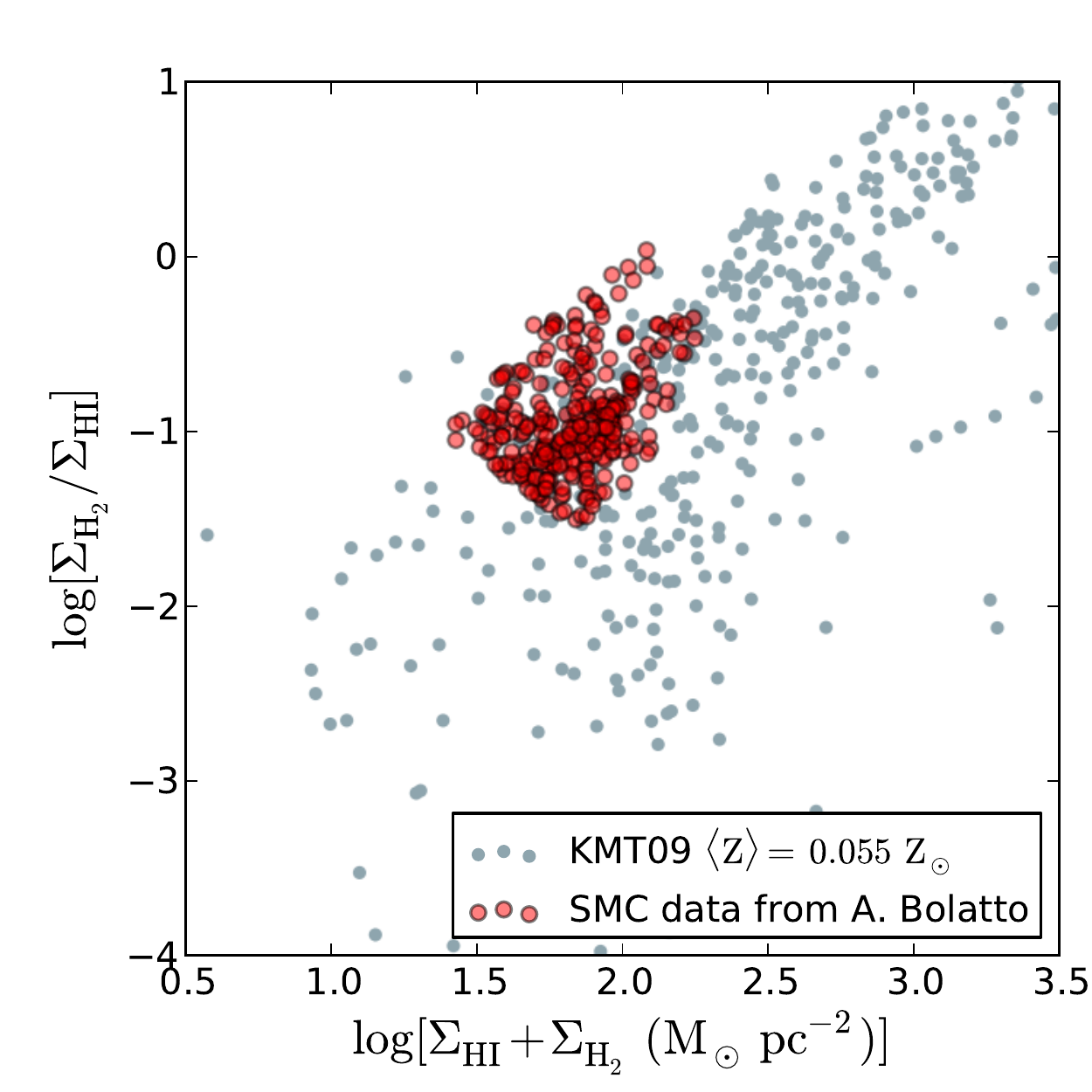}
\caption{Comparison of the \HH-to-HI ratio versus total neutral hydrogen column density ($\SigmaHI + \SigmaHH$) between
  SMC data smoothed at 200 pc \citep[from][]{bolatto_state_2011}
  and low metallicity gas in the KMT09 simulation at $z=4$ also smoothed at
  200 pc. The overlap between simulated and observational
  data indicates that our model does an adequate
  job of capturing the atomic-to-molecular transition in real dwarf
  galaxies.}
\label{fig:H2ratio_SMC}
\vspace*{0.1in}
\end{figure}

\begin{figure*}[tp]
\begin{centering}
\includegraphics[width=0.8\textwidth]{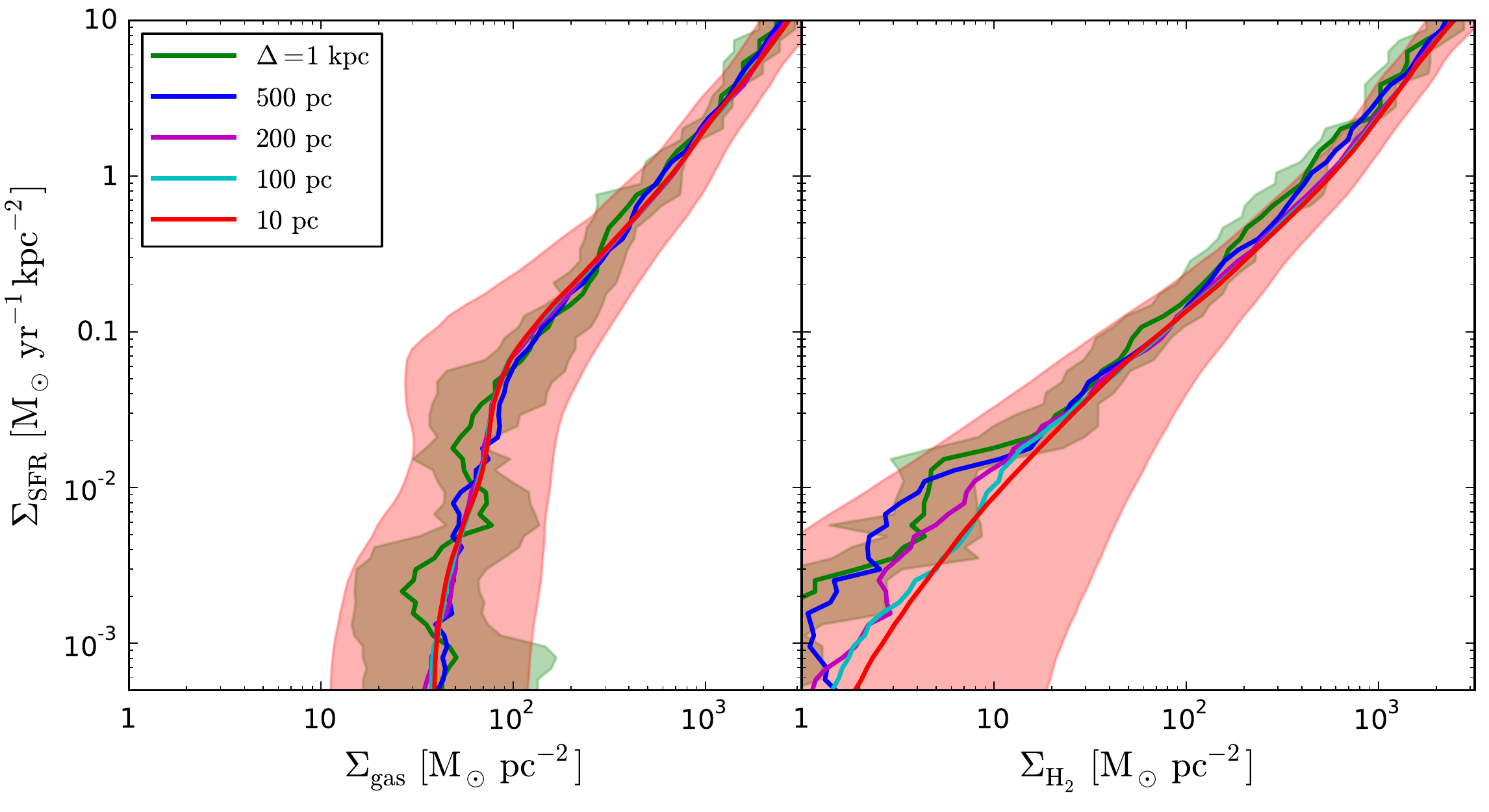}
\caption{Dependence of the total gas (\textit{left}) and
  \HH\ (\textit{right}) KS relation on the smoothing scale $\Delta$ in the
  KMT09 simulation at $z=4$. The solid lines show  the mean relations and the shaded regions cover the central 68\% scatter (for clarity we only show the scatter for $\Delta =$ 10 and 1000 pc). The mean relation does not show much dependence on averaging scale, but the scatter increases towards smaller $\Delta$, in agreement with observations \citep{schruba_scale_2010,liu_super-linear_2011}.}
\label{fig:Kennicutt_res_dependence}
\end{centering}
\vspace*{0.1in}
\end{figure*}

In local normal disk galaxies stellar velocity dispersions are
typically only 5 - 10 km s$^{-1}$, preventing stars from wandering
over more than a small fraction of the $\sim$ kpc smoothing scale in
the $\sim$ 10 Myr over which they produce significant ionizing
luminosities that may be detected in H$\alpha$ or other nebular
emission lines. Typical velocity dispersions, however, may well be
larger at higher redshift (see \citet{cresci_sins_2009} for some
empirical evidence for this), as expected if they are set by
cosmological accretion \citep{krumholz_dynamics_2010} instead of cold
disk dynamics. In fact, we find that averaged on 1 kpc scales our
galaxies have a mean 1D stellar velocity dispersion\footnote{We only
  include cells containing more than 10 star particles in the
  average.} of 44 km s$^{-1}$ at z=4. In 5\% of all cells (19 in
total) the 1D stellar velocity dispersion exceeds 100 km s$^{-1}$. At
these velocities wandering becomes more important, and as a result an
increased scatter in the KS relation should be expected at high
redshift. We have verified that the amount of scatter at low
$\SigmaGas$ increases when we reduce the spatial averaging scale
of the surface density maps (see Fig.~\ref{fig:Kennicutt_res_dependence}) or
increase the temporal SF timescale $t_\star$ \citep[see
  also]{feldmann_time_2011}.

In Figure~\ref{fig:Kennicutt_H2} we compare the KS relations for the
total, atomic, and molecular gas in the KMT09 simulation. The
\HH\ relation does not exhibit a cutoff at low column densities, since
we have not imposed any explicit density threshold in the
\HH-regulated SF prescription. The \HH\ relation is shallower than the
total gas one, and its slope is in excellent agreement with the
observational determination of the slope of the molecular gas KS
relation by \citet{genzel_study_2010}. These authors studied a
population of ``normal'' star forming galaxies at $z = 0 - 3.5$ and a
population of luminous $z \sim 0$ and $z \sim 1-3.5$ mergers
(LIRGs/ULIRGs and SMGs) and found that both had equal \HH\ KS slopes
of 1.17 with about a 1 dex higher normalization for the luminous
merger sample. The two best-fit relations from their work are shown as
dashed lines in Figure~\ref{fig:Kennicutt_H2}, and our \HH\ KS
relation lies right in between the two. The HI relation instead is
much steeper and doesn't extend much beyond $70 \msun$ pc$^{-2}$. The
near constant offset at low $\SigmaGas$ between the total gas and HI
KS relations is due to the presence of a signification amount of
ionized gas on $\sim$ kpc scales. The overall picture matches the
results reported by GK10 and is in qualitative agreement with the
empirical findings reported by \citet{bigiel_star_2008}, confirming
that the origin of the turn over in the total gas KS relation is
indeed the transition from predominantly atomic to fully molecular
gas.

\subsection{Metallicity Dependence}
\label{sec:Kennicutt_Z_dependence}

The KS cutoff occurs at somewhat higher $\SigmaGas$ in both KT07 and
KMT than in the observational data from \citet{bigiel_star_2008}. This
can be understood as a result of the lower metallicities in our
simulated galaxies. We split our 35 galaxy sample into three subsamples of different mean gas phase metallicities, $\langle \rm Z/\Zsun \rangle = $ 0.055, 0.17, and 0.32. Each subsample contains at least 10 galaxies. The highest metallicity galaxy has $\rm Z = 0.56 \, \Zsun$,
considerably below the closer to solar metallicity sample of $z=0$
field galaxies analyzed by \citet{bigiel_star_2008}. The lower the
metallicity of the gas, the higher the total gas surface density that
is required in order to provide enough LW shielding to allow the
transition to fully molecular gas. The $\SigmaGas$ scale at which the
KS cutoff occurs is thus expected to scale inversely with metallicity,
and this is exactly what Figure~\ref{fig:Kennicutt_Z} shows. The cutoff in the KS relation shifts to progressively higher $\SigmaGas$ for decreasing mean metallicity, occuring at $\gtrsim 100
\msun$ pc$^{-2}$ for the lowest metallicity case. A similar trend was previously reported by GK10. Recent
observations of star formation in the Small Magellanic Cloud (SMC)
show that the break in the total gas star formation law is indeed
shifted to higher surface density by a factor of $\Zsun / {\rm
  Z_{SMC}} \sim 5$, in precisely the manner that our models predict
\citep{bolatto_state_2011}. The higher $\SigmaGas$ cutoff in the KS
relation for low metallicity systems may also be responsible for the
observational results that $\SigmaSFR$ in damped Lyman-$\alpha$ (DLA)
systems \citep{wolfe_searching_2006} and Lyman-break galaxies
\citep{rafelski_deep_2009,rafelski_star_2011} at $z \approx 3$ appears
to lie well below the $z=0$ KS relation (see GK10 for a more in depth
exploration of this possibility).

The success of our model at capturing the physics governing the gas
phase structure in dwarf galaxies is further demonstrated by
Fig.~\ref{fig:H2ratio_SMC}. There we compare the \HH-fraction
($\SigmaHH/\Sigma_{\rm HI}$) as a function of total neutral hydrogen gas column
($\SigmaHI + \SigmaHH$) for the low metallicity galaxy sample ($\rm \langle Z \rangle = \, 0.055 \, \Zsun$) with
observational data for the SMC that has recently become
available. \citet{bolatto_state_2011} have determined a \HH\ column
density map of the SMC at $\sim 12$ pc resolution by combining
\textit{Spitzer} IR measurements and radio (ATCA and Parkers) HI
data. In Fig.~\ref{fig:H2ratio_SMC} we compared their data smoothed on
$\sim$ 200 pc scale (kindly provided by A.~Bolatto) to our $z=4$ KMT09
simulation data smoothed at the same scale. There is good overlap between simulation and observational data, but in our simulations the distribution of points extends both to higher total gas columns and to lower \HH-fractions than probed by \citet{bolatto_state_2011}.

\subsection{Smoothing Scale Dependence}
\label{sec:Kennicutt_res_dependence}

\begin{figure*}[tp]
\begin{centering}
\includegraphics[width=0.4\textwidth]{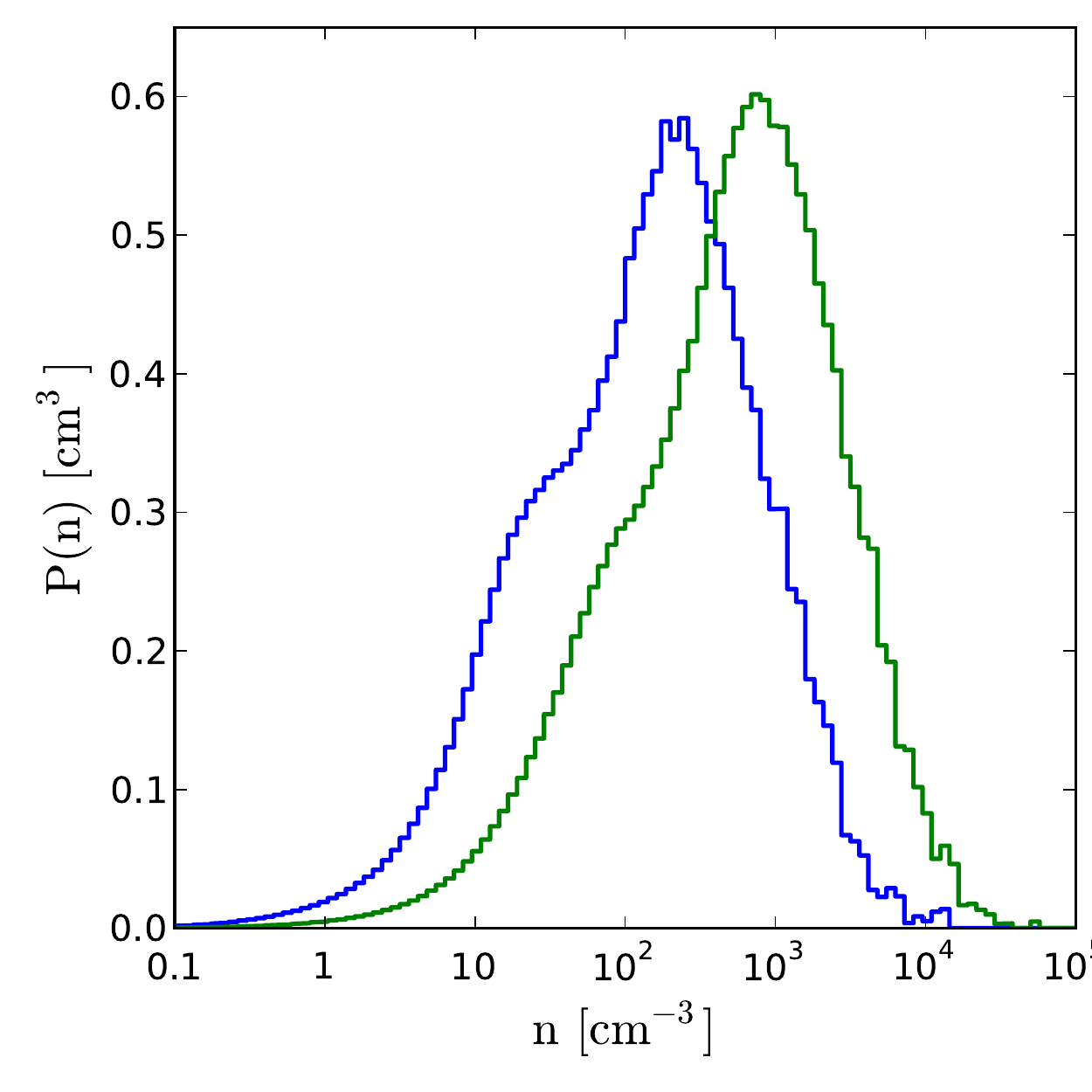}
\includegraphics[width=0.4\textwidth]{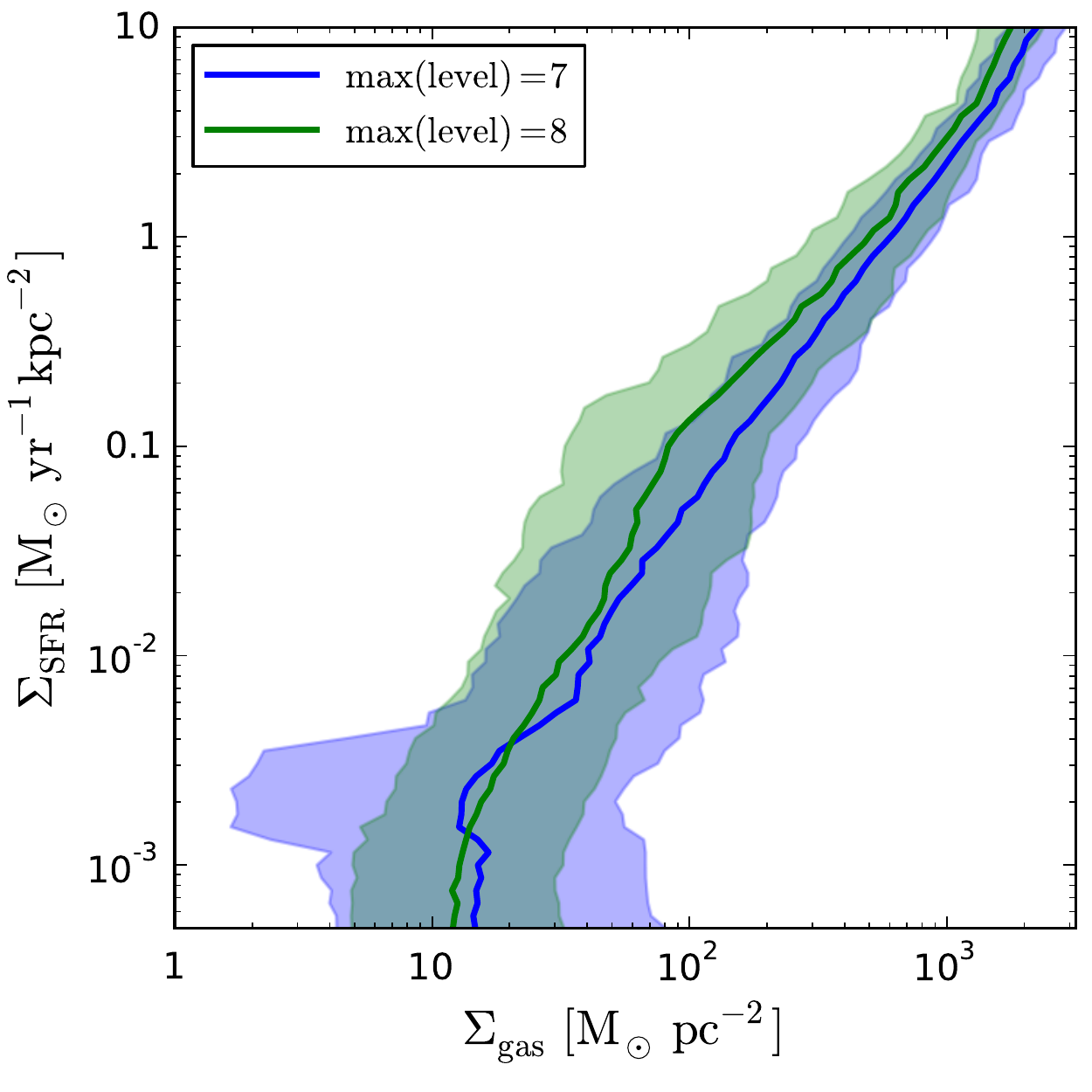}
\caption{\textit{Left:} Probability distribution functions of the
  proper gas number density in the KMT09 and KMT09\_L8 simulations at
  $z=6$ for cells at the maximum refinement level ($l_{\rm max}=7$ and
  8, respectively). \textit{Right:} The corresponding total gas KS
  relation.  The additional refinement level allows gas in KMT09\_L8
  to reach higher densities, which in turn results in greater SFR
  densities, and an increase in the normalization of the KS relation.}
\label{fig:KS_resolution}
\end{centering}
\vspace*{0.1in}
\end{figure*}

In Figure~\ref{fig:Kennicutt_res_dependence} we show the dependence of the
total gas and \HH\ KS relations on the spatial scale over which the
data is smoothed. In addition to our fiducial scale of 200 pc, we present the relations for up to twenty times finer and five times coarser smoothing. This range of smoothing scales roughly mimics
the variations in the angular resolution (i.e. beam size) of the radio
observations used to establish the observational KS relation, ranging
from spatially resolved measurements of nearby spiral galaxies with
sub-kpc resolution \citep{kennicutt_star_2007,bigiel_star_2008} to
high redshift observations in which a large fraction of the galaxy is contained in
a single beam \citep{kennicutt_global_1998}. We do not see much evidence for a smooth scale dependence, in either total gas or \HH\ KS relations. The scatter in the relations, however, increases for smaller smoothing scales, in agreement with observations
\citep{schruba_scale_2010,liu_super-linear_2011}.


\subsection{Resolution Dependence}
\label{sec:KS_resolution}

With only seven levels of adaptive mesh refinement, our simulations
are unable to resolve the true Jeans length of the cold, molecular gas
in star forming galaxies. As discussed in
\S~\ref{sec:simulations}, we resort to an artificial minimum
pressure support in order to stabilize gas cells at the highest
refinement level against artificial fragmentation. This has the
undesirable consequence of making the results of our simulations
somewhat dependent on resolution, since additional levels of
refinement will allow gas to collapse further and reach higher
densities, until the resolution becomes adequate to resolve the true
Jeans length. Unfortunately, additional refinement levels come at a
computational cost. To run the KMT09\_L8 simulation, a clone of KMT09
with one additional refinement level ($l_{\rm max}=8$), down to $z=6$
took about 3 times as long as the KMT09 run. Note that virtually every
$l=7$ KMT09 grid was at least partially further refined in KMT09\_L8.

In the left panel of Figure~\ref{fig:KS_resolution} we show a comparison
of the distribution functions of number density in the maximally
refined grid cells in the KMT09 and KMT09\_L8 simulations at
$z=6$. The additional refinement level has allowed gas to collapse to
higher densities. The mean density at $l=7$ is 427 cm$^{-3}$ in the
KMT09 simulation, but 1420 cm$^{-3}$ at $l=8$ in KMT09\_L8. Higher
densities will lead to larger a SFR and an increase in
$\SigmaSFR$. $\SigmaGas$ smoothed on $\sim$ kpc scales, however, will
be unaffected, since it depends only on the \textit{total mass}
enclosed in a given kpc scale column, not the local density. It is
not surprising, then, that the simulated KS relation (right panel of
Fig.~\ref{fig:KS_resolution}) shows that the KMT09\_L8 KS relation has
somewhat higher amplitude than in KMT09.

It is important to note that, despite its higher resolution, model
KMT09\_L8 is not necessarily more realistic than KMT09, because it
lacks the physics needed to properly model molecular clouds at the
increased resolution. In the absence of feedback mechanisms other than
supernovae, increasing resolution allows the gas to collapse to
ever-higher density, so that the bulk of the molecular gas will always
reside near the resolution limit. However, this behavior is not
realistic. In observed nearby galaxies, the bulk of the molecular
clouds exist at densities of a few hundred cm$^{-3}$ rather than a few
thousand cm$^{-3}$ \citep[i.e.\ closer to the mean in KMT09 than
  KMT09\_L8;][]{blitz_giant_1993}, and molecular cloud properties are
independent of galactic environment, strongly suggesting internal
regulation \citep{bolatto_resolved_2008}. Possible mechanisms to
provide this regulation include ionized gas pressure
\citep{matzner_role_2002,krumholz_global_2006,goldbaum_global_2011},
protostellar winds
\citep{nakamura_protostellar_2007,wang_outflow_2010}, and radiation
pressure
\citep{krumholz_dynamics_2009,murray_disruption_2010,fall_stellar_2010,hopkins_self-regulated_2011},
none of which are included in our simulations. Thus we regard the KS
law we obtain from KMT09 as at least as reliable as the one we obtain
from KMT09\_L8. Moreover, this comparison yields an important caution:
increasing resolution without a corresponding increase in physics does
not necessarily yield a better result.

\begin{figure*}[htp]
\begin{centering}
\includegraphics[width=0.8\textwidth]{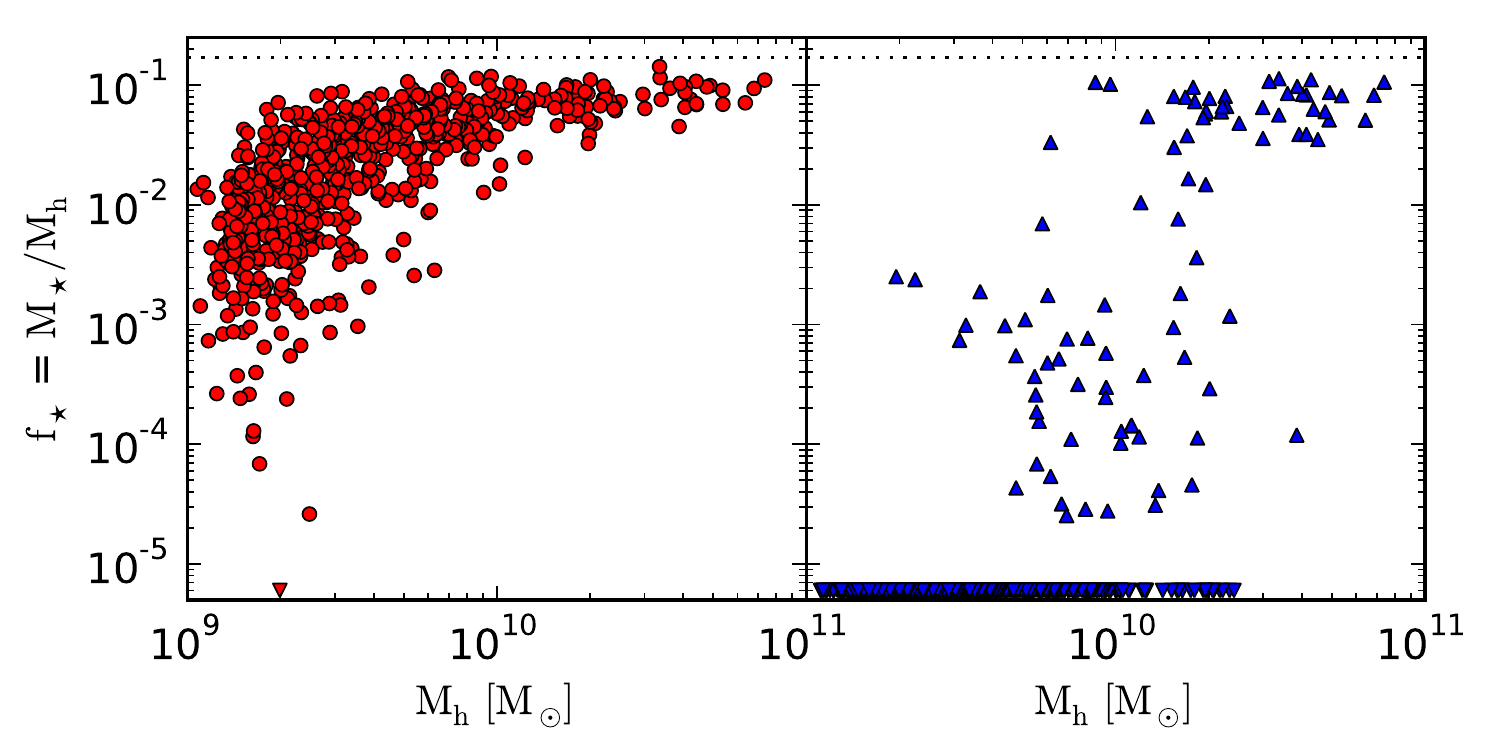}
\caption{Stellar mass fraction $\fstar$ vs. total halo mass in a
  simulation with standard SF (KT07, \textit{left}) and with
  \HH-regulated SF (KMT09\_ZFz10, \textit{right}) at $z=4$. The dotted
  horizontal line indicates an $\fstar$ equal to the cosmic baryon
  fraction $\Omega_{\rm b}/\Omega_{\rm M}$, i.e. a 100\% gas to star
  conversion efficiency.}
\label{fig:fstar}
\end{centering}
\vspace*{0.1in}
\end{figure*}

\section{Stellar Mass Fraction}
\label{sec:fstar}

\begin{figure*}[tp]
\includegraphics[width=\textwidth]{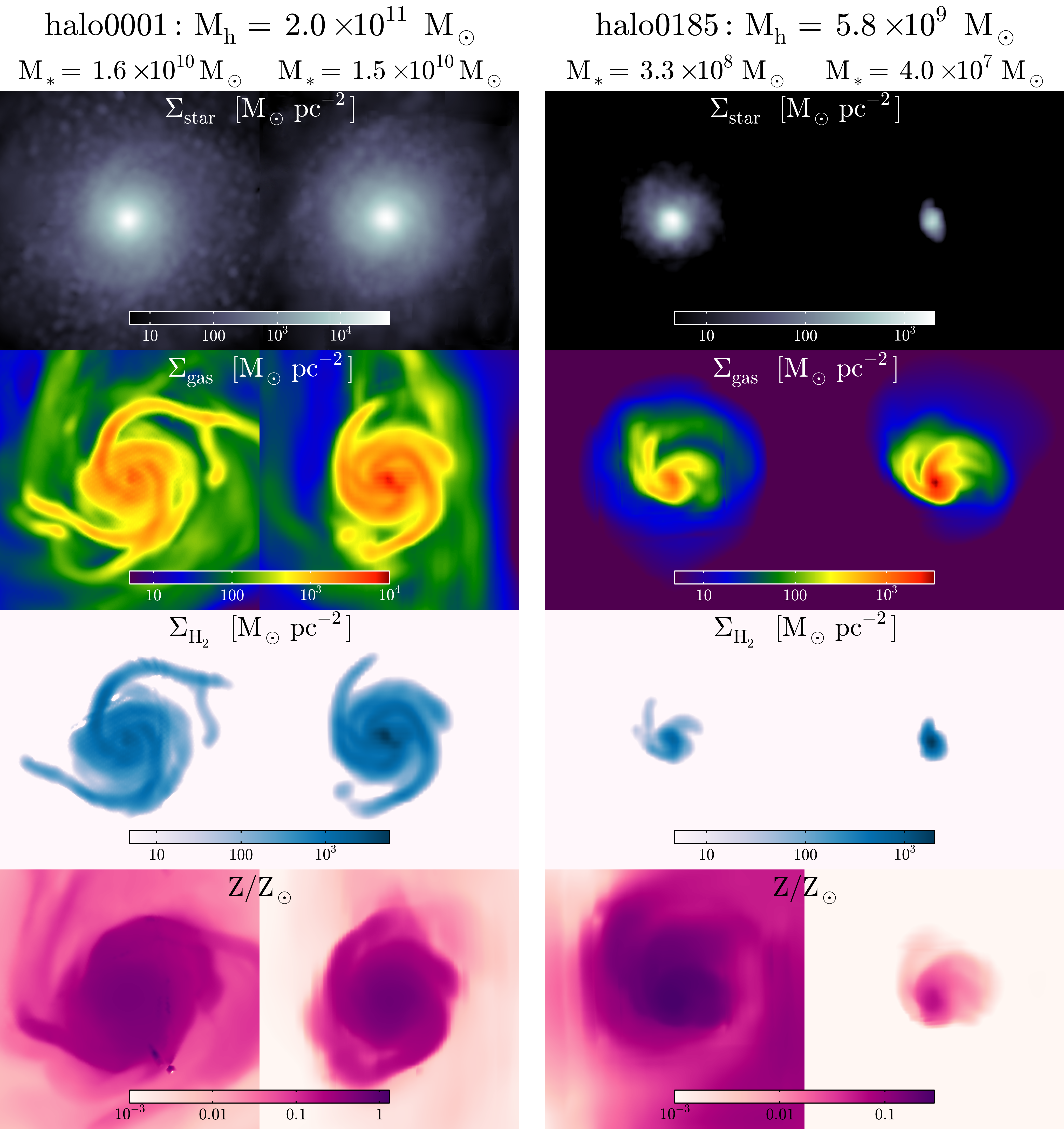}
\caption{A comparison of the baryonic structure of two identical halos, one with high total mass (halo0001, left panel) and one with low total mass (halo0185, right panel), between the KT07 (left columns) and KMT09\_ZFz10 (right columns) simulations at $z=4$. From top to bottom, we show surface densities of stellar mass, total gas, and \HH, and a density-weighted projection of metallicity, in a $5 \times 5$ kpc region centered on the galaxies. These are merely two representative halos, and many more like them exist in our simulations. The stellar content is greatly suppressed in the KMT09 low mass halo.}
\label{fig:multi_panel_comparison}
\vspace*{0.1in}
\end{figure*}

We now turn to the effects of an \HH-regulated SF prescription on the
stellar content of the dark matter halos in our simulations. In
Figure~\ref{fig:fstar} we show plots of the stellar mass fraction
$\fstar = M_\star / M_{\rm h}$ against the total halo mass $\Mhalo =
\Mdm + \Mgas + \Mstar$ for simulations with a standard (KT07) and
\HH-regulated (KMT09\_ZFz10\footnote{We now focus on KMT09\_ZFz10,
  because the slightly earlier metallicity floor ($z=10$, instead of
  $z=9$ in KMT09) results in a more gradual suppression of $\langle
  \fstar \rangle$, see \S~\ref{sec:fstar_Zfloor}.}) SF
prescription. In the normal SF case (left panel), halos with total
masses as low as $10^9 \msun$ have been able to form a substantial
stellar content, with values of $\fstar$ not dropping much below
1\%. This is problematic in view of the observational dearth of such
dwarf galaxies in the local universe. Given the high volume density of
${\rm M} > 10^9 \msun$ dark matter halos predicted by $\Lambda$CDM
structure formation, such a high star formation efficiency would
vastly overproduce the faint end of the field dwarf galaxy luminosity
function and the abundance of faint Local Group dwarf galaxies.

\begin{figure}[htp]
\includegraphics[width=\columnwidth]{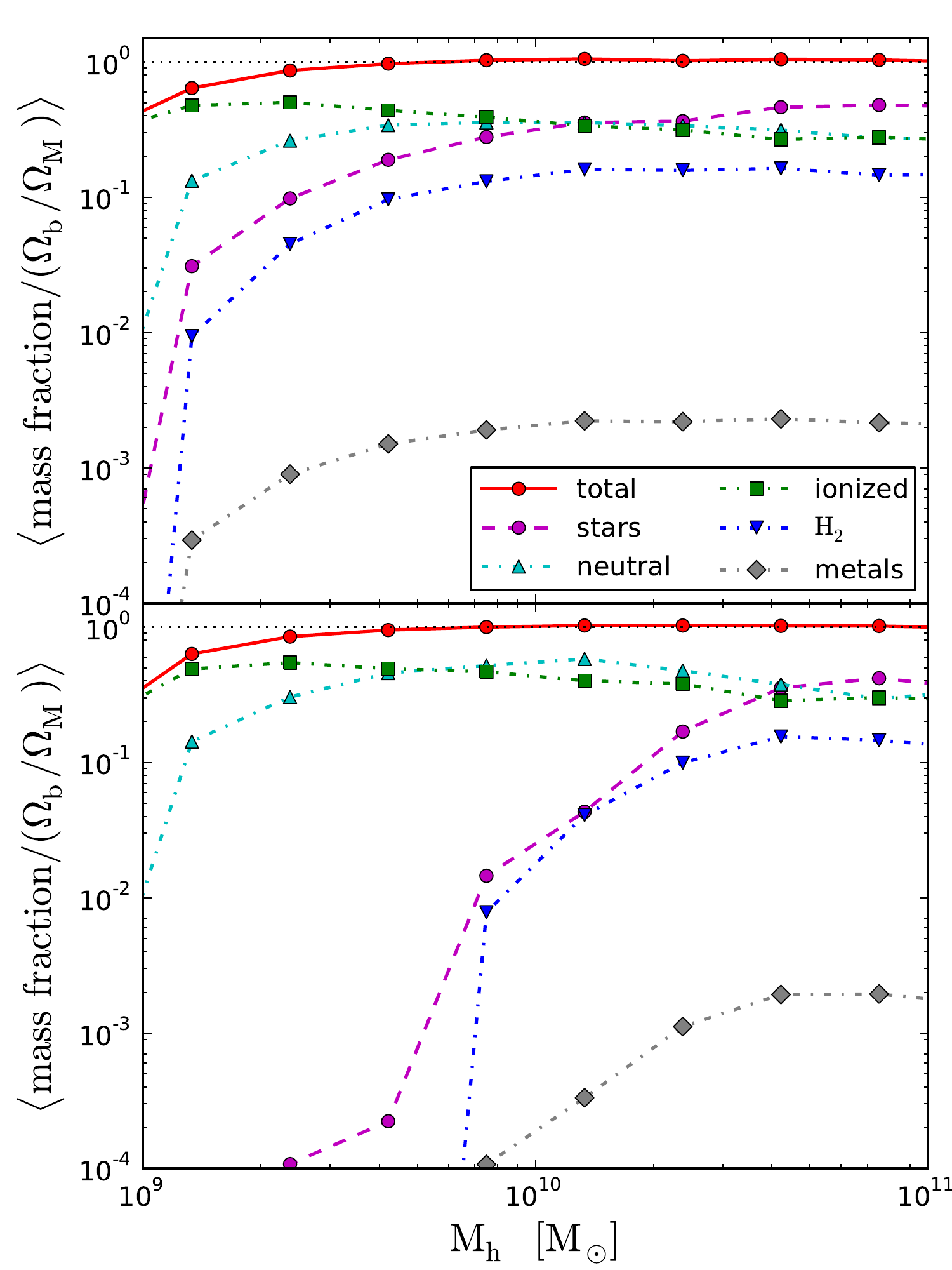}
\caption{The baryonic content of halos versus their total mass, for
  the KT07 (\textit{top panel}) and KMT09\_ZFz10 (\textit{bottom
    panel}) simulations at $z=4$. We show mean mass fractions
  normalized to the cosmic baryon fraction ($\Omega_b/\Omega_M$) in
  $\Mhalo$-bins of width 0.25 dex for the total baryonic content
  (stars + all gas + metals; solid lines with circles), stars (dashed
  with circles), neutral gas (HI + HeI + \HH; dot-dashed with upward
  triangles), ionized gas (HII + HeII + HeIII; dot-dashed with
  squares), \HH\ (dot-dashed with downward triangles), and metals
  (dot-dashed with diamonds). Note that although an \HH\ curve is
  plotted in the KT07 panel, \HH\ has no effect in that simulation.}
\label{fig:halo_baryon_content}
\vspace*{0.1in}
\end{figure}

\begin{figure*}[tp]
\begin{centering}
\includegraphics[width=0.9\textwidth]{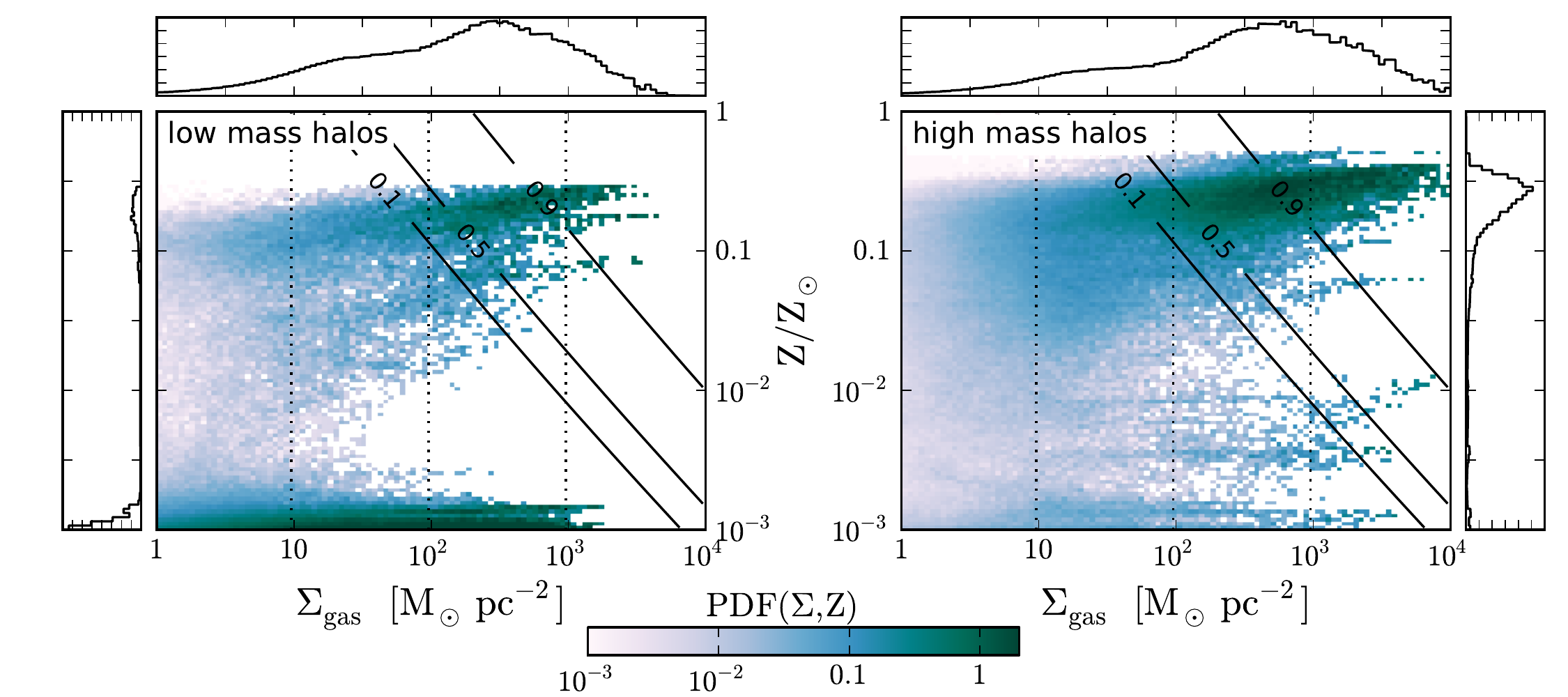}
\caption{2D phase diagrams of $\SigmaGas$ vs. Z determined at $l_{\rm
    max}=7$ in the KMT09 simulation at $z=5$, for low mass (M$<10^{10}
  \msun$, \textit{left panel}) and high mass (M$>10^{10} \msun$,
  \textit{right panel}) halos. Contours of constant $\fHH$ (at 0.1,
  0.5, and 0.9) are plotted with solid lines. For reference we show
  with dotted lines the column densities corresponding to the 3D
  density thresholds employed in the KT07 (50 cm$^{-3}$), KT07\_low (5
  cm$^{-3}$), and KT07\_high (500 cm$^{-3}$) simulations. The majority
  of gas in low mass halos has very little or no \HH, and this appears
  to be primarily due to lower metallicities, not lower column
  densities.}
\label{fig:SigmaZphase}
\end{centering}
\vspace*{0.1in}
\end{figure*}

In Via Lactea II, a collisionless simulation of the formation of a
Milky Way's dark matter halo \citep{diemand_clumps_2008}, for example,
there are $\sim 100$ halos with an identifiable $z=0$ remnant within
the host halo's virial volume that had a mass exceeding $10^9 \msun$
at some point in their evolution. Yet of the Milky Way dwarf satellite
galaxies, probably only the two Magellanic Clouds, Fornax, and Leo I
(and possibly the progenitor of the Sagittarius dwarf) have a stellar
mass greater than $10^7 \msun$. The vast majority of such dark matter
halos must thus have SF efficiencies well below 1\%
\citep{rashkov_assembly_2012}. The same is true of faint field galaxies, not
just satellites. Comparisons of the local universe galaxy luminosity
function from SDSS with the predicted DM field halo mass function show
that at $z<1$ the stellar mass fraction must have a maximum of $\sim
0.01 - 0.03$ around $10^{12} \msun$, decreasing sharply towards lower
masses \citep{zheng_galaxy_2007, conroy_connecting_2009, guo_how_2010,
  moster_constraints_2010, behroozi_comprehensive_2010}.

The most commonly invoked mechanisms to lower the stellar mass
fraction in these low mass galaxies are (i) the prevention of gas
cooling due to the meta-galactic UV background
\citep{efstathiou_suppressing_1992,kauffmann_formation_1993,bullock_reionization_2000}
and (ii) stellar feedback from massive winds and supernovae
\citep{larson_effects_1974,kauffmann_formation_1993,benson_effects_2002}. Our
simulations include a model for the UV background, via a spatially
uniform, optically thin radiation field that accounts for the UV
emissivity from both quasars and galaxies\footnote{The galactic
  contribution dominates the UV background at this early epoch.}
\citep{haardt_modelling_2001}. By itself this feedback does not appear
to be able to suppress star formation in $10^9 - 10^{10} \msun$ halos
at $z>4$ in our simulations. As discussed in \S~\ref{sec:feedback}, we
do include a supernova feedback prescription in our simulation, but in
a form known to be insufficiently strong to effect the necessary
reduction in SF efficiency or to reproduce observed galactic
outflows. This is exemplified by the high stellar mass fractions
($\fstar \approx 0.06$, a factor $5 - 10$ higher than what is allowed
by observational constraints at much lower redshift, $z \lesssim
  1$) even in our most massive halos ($10^{10} - 10^{11} \msun$)
\citep[see also][]{avila-reese_specific_2011}.

It's reasonable to expect a similar, or even greater, suppression of
$\fstar$ for lower mass halos, whose shallower potential wells should
make it easier for winds and supernovae to expel their gas. However,
\citet{font_new_2011} find that in order to simultaneously match the
luminosity function of Local Group dwarf galaxies and their observed
mass-metallicity relation, the strength of stellar feedback must
saturate in halos with $V_{\rm max} \lesssim 65$ km s$^{-1}$ ($\approx
5 \times 10^{10} \msun$). Furthermore, a physically realistic modeling
of stellar feedback processes in cosmological numerical simulations is
beyond current computational capabilities, and it remains unclear how
well the various sub-grid physics implementations discussed in the
literature actually capture the true nature of this feedback. Although
we don't doubt that stellar feedback processes play an important role
in regulating star formation in low mass halos, we now show that
regulating SF by the \HH\ abundance may play an equally important role
in lowering stellar mass fractions in low mass ($<10^{10} \msun$)
halos, as previously suggested by G09, GK10.

The right panel of Figure~\ref{fig:fstar} shows that our \HH-regulated
SF prescription introduces a threshold below which the stellar mass
content of halos is strongly suppressed. At halo masses greater than
$10^{10} \msun$, the star formation efficiency is only mildly reduced,
dropping from $\langle \fstar \rangle = 0.062$ to 0.035. Between $5
\times 10^9 \msun$ and $10^{10} \msun$, only a small fraction of halos
(about 10\%) has been able to form stars and only with reduced SF
efficiency, and at even lower halo masses star formation has been
almost completely suppressed. We visually demonstrate this suppression in Figure~\ref{fig:multi_panel_comparison}, where we show direct comparisons of the baryonic structure ($\Sigma_*$, $\SigmaGas$, $\SigmaHH$, and density-weighted metallicity) between the KT07 and KMT09\_ZFz10 simulations for two representative halos; one high mass halo ($\Mhalo = 2.0 \times 10^{11} \msun$) in which \HH-regulation has not had a big effect, and one low mass halo ($\Mhalo = 5.8 \times 10^{9} \msun$), in which the stellar mass fraction has been suppressed by nearly one order of magnitude from $\fstar =$ 0.057 to 0.0069. Many more halos like these exist in our simulations.

Figure~\ref{fig:halo_baryon_content} shows the baryonic content of halos
in KT07 and KMT09\_ZFz10 as a function of their mass. We plot mean mass
fractions in $\Mhalo$-bins of width 0.25 dex of the total baryonic
content, stars, neutral gas (HI + HeI + \HH), ionized gas (HII + HeII
+ HeIII), \HH, and metals. The total baryonic mass fraction remains
equal to the cosmic mean down to $\sim 2 \times 10^9 \msun$. The
slight drop in the total baryonic and neutral gas fractions at lower
halo masses can be attributed to the meta-galactic UV background,
which is able to ionize and heat most of the gas, preventing it from
falling into the halo. The mass fractions of \HH, metals, and stars
roughly trace each other. In KT07, the stellar mass fraction decreases
gradually from $\sim 0.4$ of the cosmic baryon fraction at $\Mhalo =
10^{11} \msun$ to $0.25$ at $10^{10} \msun$, before the combined
actions of UV background and (weak) supernova feedback further reduce
it to $\sim 0.01$ at $10^9 \msun$. In KMT09\_ZFz10, on the other hand, the
stellar mass fraction is suppressed already at much higher halo
masses, dropping below 0.1 of the cosmic mean at $10^{10} \msun$ and
cutting off completely at $4 \times 10^9 \msun$. Note that because
we have included halos with $\Mstar=0$ in the calculation of the mean
$\fstar$ it should not be viewed as a typical value
for any individual halo, but as a population average. With
\HH-regulation the SF is quenched in low mass halos without heating
and removing much of the gas, contrary to the effects of efficient
supernova feedback.

\begin{figure}[tp]
\includegraphics[width=\columnwidth]{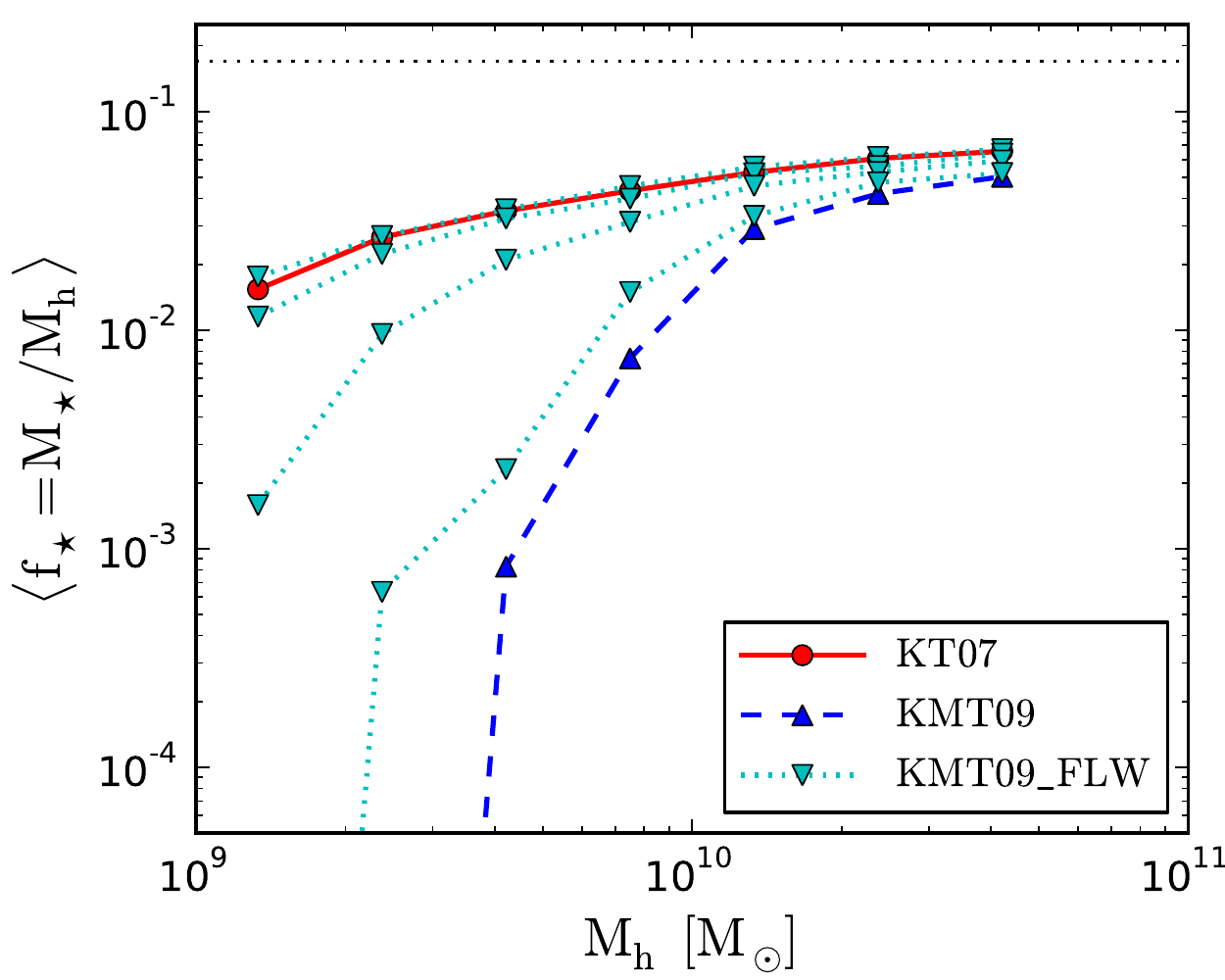}
\caption{$\langle \fstar \rangle$ vs. M$_{\rm h}$ at $z=5$ for the
  four KMT09 simulations without the two phase equilibrium assumption
  and a spatially uniform Lyman-Werner radiation field with intensity
  equal to 1, 10, 100, 1000 $\times$ the mean Milky Way value (dotted
  lines with downward triangles; increasing LW intensity from top to
  bottom). We plot the mean of $\fstar$ in M$_h$-bins of width 0.25
  dex. The means include halos with $\fstar=0$. For comparison we also
  show the original KT07 (solid line with red circles) and two-phase
  equilibrium KMT09 results (dashed line with upward triangles). Note
  that for the KMT09\_FLW simulations (but not for KMT09) we apply a
  ``clumping factor'' of 30 to the \HH\ formation rate (see text for
  details).}
\label{fig:fstar_FLW}
\vspace*{0.1in}
\end{figure}

Why should a locally \HH-regulated SF prescription be sensitive to the
total halo mass? Recall that with the two-phase equilibrium assumption
(KMT09 model), the local \HH\ abundance is completely determined by
the HI column density and the metallicity of the gas. A reduced SF
efficiency in low mass halos could thus be due to either lower column
densities or to lower metallicities than in more massive halos, or to
a combination of the two.

Figure~\ref{fig:SigmaZphase} reveals that lower metallicities are
primarily responsible. The figure shows two-dimensional volumetric
probability density functions (phase diagrams) of total gas column
density $\SigmaGas$ and metallicity ${\rm Z}$, for $l=7$ grid cells in
M$<10^{10} \msun$ (left panel) and M$>10^{10} \msun$ halos (right
panel) halos. The color scale indicates the volume fraction at a given
$(\SigmaGas, {\rm Z})$, and the diagonal lines represent contours of
constant $\fHH$, according to the KMT09 prescription. The
one-dimensional distributions at the top and sides of the figure show
that while both low and high mass halos have comparable column density
distributions, the metallicity distribution peaks at ${\rm Z}/\Zsun
\approx 0.25$ for the high mass halos, but at ${\rm Z}/\Zsun=10^{-3}$
for the low mass halos. Recall that we imposed a ${\rm
  Z}/\Zsun=10^{-3}$ metallicity floor at $z=9$, so this indicates that
low mass halos typically remain unenriched. This lack of metals in low
mass halos is responsible for a reduced \HH\ abundance, and hence a
suppressed stellar mass fraction.

As we discuss in the remainder of this section, a number of factors,
both of physical nature and pertaining to the limitations of our
simulations, affect the mass scale and the degree of the $\fstar$
suppression due to \HH-regulated SF. A complete suppression of SF in
halos with masses below $5 \times 10^9 \msun$ is too strong of an
effect to reproduce the luminosity function of Local Group satellites,
many of which are consistent with having formed in $\lesssim 10^9 \msun$
halos \citep{madau_dark_2008}. In this work, we demonstrate that
\HH-regulated SF can have important consequences for the star
formation efficiency in low mass halos, but openly acknowledge that
the true nature of this suppression will depend on details of the star
formation processes, stellar feedback, and metal enrichment that are
not captured with sufficient fidelity in our simulations.

\subsection{Dependence on Two-Phase Equilibrium}

The assumption of two-phase equilibrium between a Cold Neutral Medium
(CNM), hosting molecular clouds and SF, and the surrounding Warm
Neutral Medium (WNM) is not uncontroversial. As discussed in
\citet{krumholz_atomic--molecular_2009}, shocks from supernova
explosions or supersonic ISM turbulence, for example, likely
temporarily drive the surrounding gas out of pressure balance, in
which case the typical gas density could differ significantly from the
value estimated from two-phase equilibrium. The time to re-establish
equilibrium between a CNM and WNM can be comparable (few Myr,
\citet{wolfire_neutral_2003}) to the shock recurrence time scale, and
hence it is prudent to consider the effects of the KMT09 model without
the two-phase equilibrium assumption.

In this case the molecular hydrogen fraction is no longer independent
of the \HH-dissociating LW background, and absent a full radiation
transfer treatment (G09) we need to externally specify its
intensity. For computational ease we have considered here only the
case of a constant, spatially uniform LW intensity. Of course it would
be preferable to tie the LW intensity to the local SFR, averaged over
$\sim$ kpc scales and tens of Myr, but we defer the investigation of
this more realistic treatment to future work.

In Figure~\ref{fig:fstar_FLW} we show $\langle \fstar \rangle$
vs. M$_{\rm h}$ at $z=5$ for simulations KMT09\_FLW\{1,10,100,1000\},
with values of the LW intensity ranging from 1 to 1000 times the $z=0$
Milky Way value of $7.5 \times 10^{-4}$ LW photons cm$^{-3}$
\citep{draine_photoelectric_1978}. For clarity we have plotted the
mean values of $f_\star$ in bins of $\Mhalo$ of width 0.25 dex. The
means include halos with $f_\star=0$, which implies that these values
of $\langle \fstar \rangle$ should thus be viewed as population
averages, rather than representative values for any individual halo.
The results from the standard SF (KT07) and the two-phase equilibrium
(KMT09) simulations are included for comparison. The lowest LW
intensity case (KMT09\_FLW1) is almost indistinguishable from the
standard SF model. In this case the dissociating flux is not strong
enough to affect the \HH\ abundance even in the lowest mass halos. As
the intensity of the LW background is increased, the halo mass at
which \HH\ formation begins to be suppressed increases in
proportion. At 1000 times the Milky Way's LW intensity (lowest dotted line), the suppression mass almost reaches the value in the two-phase
equilibrium model. Recall that the KMT09\_FLW simulations were run
with an effective sub-grid clumping factor of 30
($(\sigma_{d,-21}/\mathcal{R}_{-16.5})=1/30$), and the resulting
enhanced \HH\ recombination rate is partially responsible for the
reduced SF suppression compared to the two phase equilibrium
simulation. However, while the absolute strength of the SF suppression
in the simulations without two-phase equilibrium is sensitive to the
value of the clumping factor, the trend with the intensity of the LW
background is not.

\subsection{Resolution Dependence}
\label{sec:fstar_resolution}

\begin{figure}[tp]
\includegraphics[width=\columnwidth]{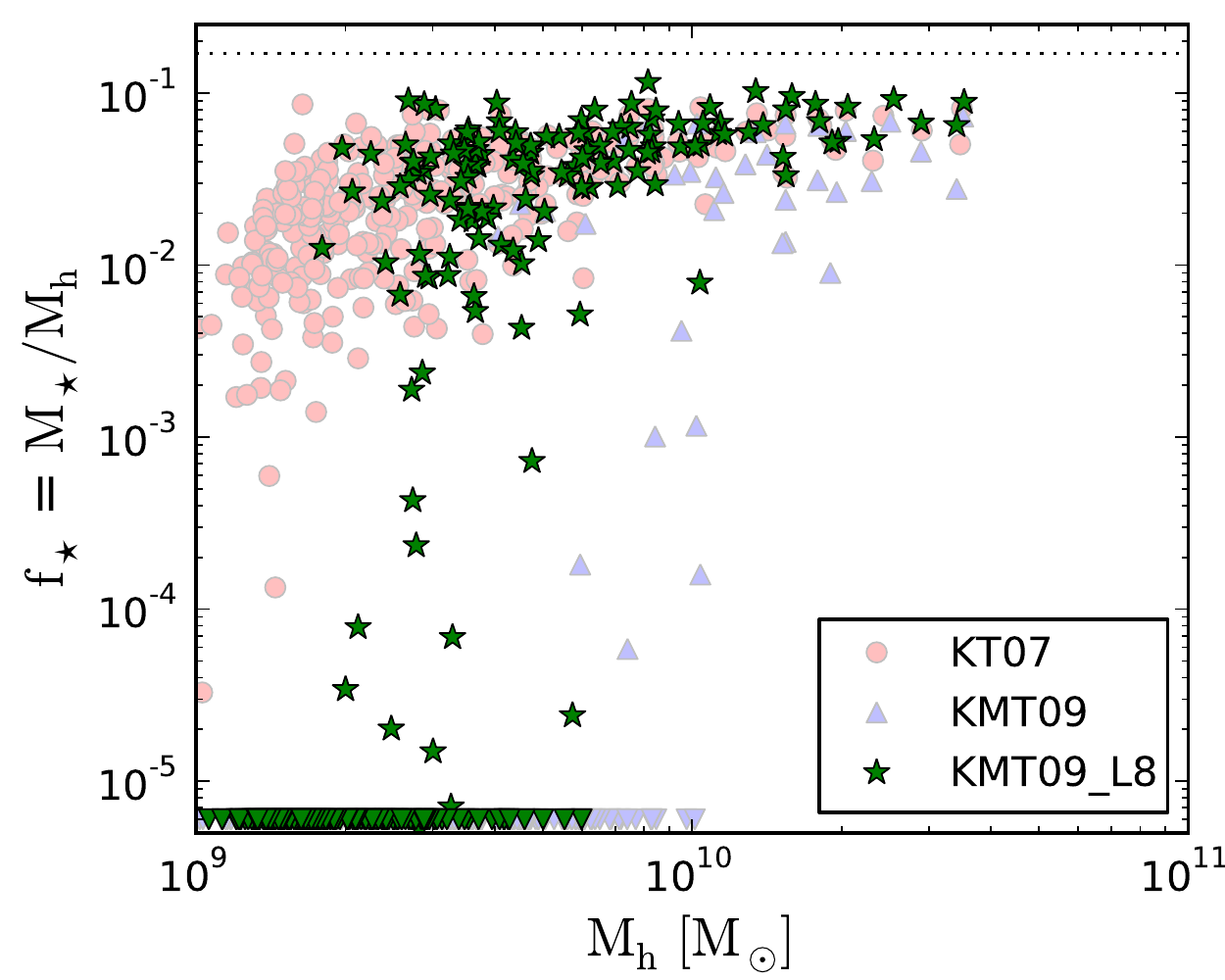}
\caption{$\fstar$ vs. halo mass (at $z=6$) for the KMT09\_L8
  simulation, which has one additional refinement level compared to
  KMT09 and KT07. The halo mass scale at which star formation is
  suppressed due to the inability of gas to become molecular is lower
  than in the KMT09 simulation. The true value of this suppression
  mass scale in nature will depend on the strength of stellar
  feedback, metal mixing, and other processes not resolved in our
  simulations.}
\label{fig:fstar_resolution}
\vspace*{0.1in}
\end{figure}

In Figure~\ref{fig:fstar_resolution} we compare $\fstar$ vs. halo mass for
the KMT09\_L8 simulation to the results from KT07 and KMT09. While
there still is a suppression of star formation in low mass halos
compared to KT07, the scale at which this suppression occurs has
shifted to lower masses, around $3 \times 10^9 \msun$. Considering our
earlier finding (\S~\ref{sec:KS_resolution}) that the additional
resolution increases the SF rates in KMT09\_L8, it is perhaps not
surprising that the $\fstar$-suppression mass scale also exhibits some
resolution dependence. In this case the sensitivity to the maximum
resolution arises because the increase in density afforded by the
additional refinement level exceeds the factor of two reduction in
grid cell width, thus leading to a net increase in column densities on
the finest grid cells. The enhanced LW shielding eases the transition
to the molecular phase, and allows star formation to occur in lower
mass halos.

The exact value of this transition will depend on the details of the
star formation and stellar feedback processes on scales below our
current resolution limit, as we discussed in
\S~\ref{sec:KS_resolution}. However, our fiducial $l_{\rm max}=7$
grid resolution of $76.3 \times 5/(1+z)$ proper parsec is comparable
to the observed size of giant atomic-molecular cloud complexes, and
our $l=7$ grid cell averaged densities (Fig.~\ref{fig:KS_resolution},
left panel) are in good agreement with observational estimates of
their average densities \citep{blitz_giant_1993}. We thus believe that
the resolution of our fiducial simulations ($l_{\rm max} =7$) is well
matched to the problem we are studying.

\subsection{Metallicity Floor Dependence}
\label{sec:fstar_Zfloor}

\begin{figure}[tp]
\includegraphics[width=\columnwidth]{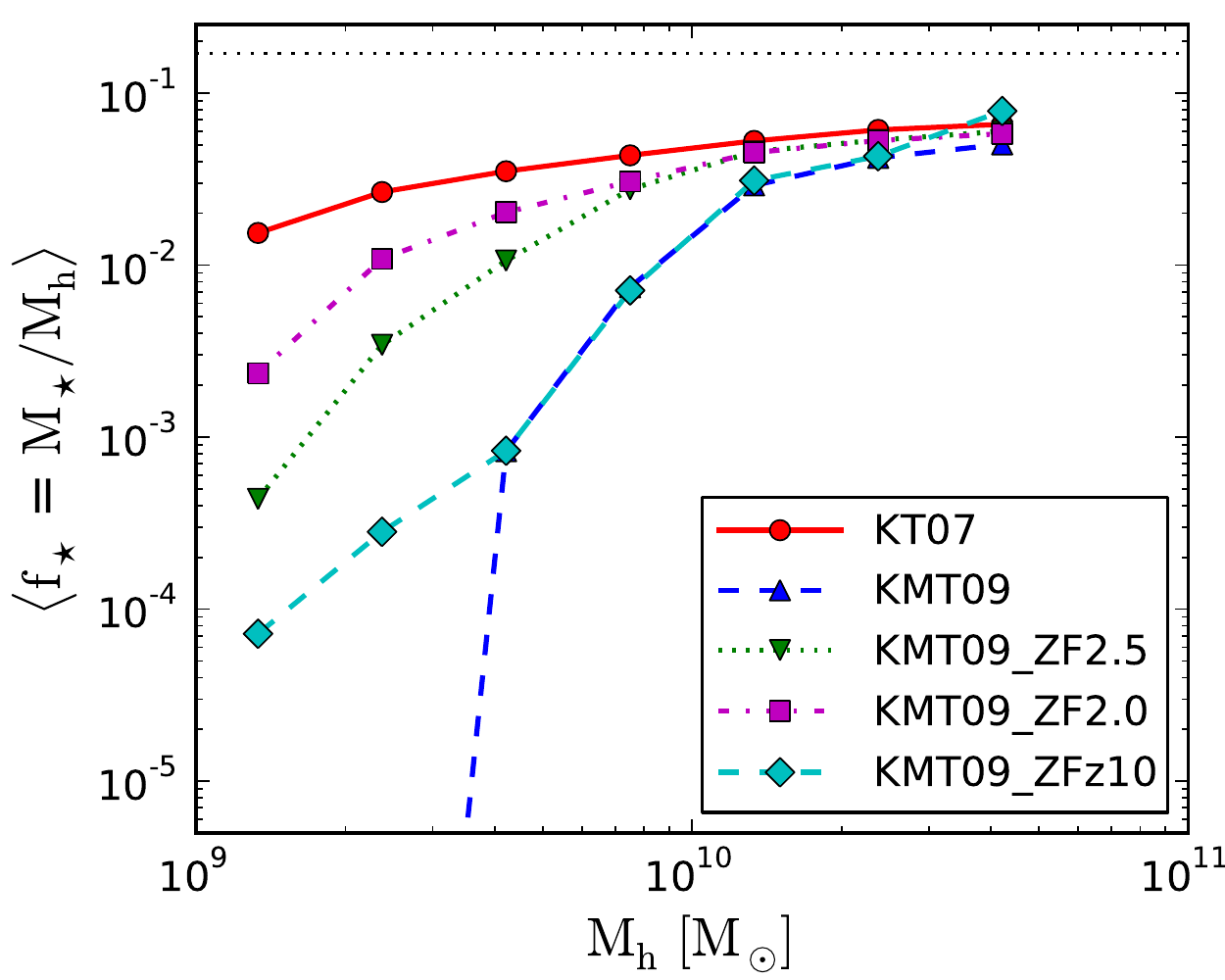}
\caption{The dependence of $\langle \fstar(\Mhalo) \rangle$ at $z=6$ on
  the metallicity floor. We plot the mean of $f_\star$ in
  $\Mhalo$-bins of width 0.25 dex at $z=6$ for KT07 (solid lines with
  circles) and three versions of KMT09 with increasing values of the
  $z=9$ metallicity floor, $[\Zfloor] = -3.0$ (dashed with upward
  triangles), $-2.5$ (dotted with downward triangles), and $-2.0$
  (dot-dashed with squares), and one case with $[\Zfloor] = -3.0$
  applied at $z=10$ (dashed with diamonds). The means include halos
  with $f_\star=0$. Note that a simulation with
  $\log_{10}(\Zfloor/\Zsun) = -4.0$ at $z=9$ produced no stars at
  all.}
\label{fig:fstar_Zfloor}
\vspace*{0.1in}
\end{figure}

In Figure~\ref{fig:fstar_Zfloor} we show how the $f_\star$ suppression
depends on the amplitude and time of the metallicity floor that we
impose early in the simulations to mimic the enrichment from Pop.III
supernovae. Higher values of $[\Zfloor]$ and an earlier time allow SF
to occur in lower mass halos. Note that the KMT09\_ZF4.0 simulation
($[\Zfloor]=-4.0$) produced no stars at all. Imposing the metallicity
floor at $z=10$ instead of $z=9$ has the particularly interesting
result of allowing a small amount of star formation in very low mass
halos ($\Mhalo < 5 \times 10^9 \msun$), which may help to alleviate
the tension with the faint-end of the Local Group satellite galaxy
luminosity function.

For a given $\Zfloor$, a halo's ability to form stars is determined by
the highest column densities that its gas can condense to prior to the
enrichment from the first star particles in our simulation.  Gas in
more massive halos is able to reach higher column densities, owing to
their earlier collapse times and deeper potentials. Below some halo
mass scale, gas will not be able to reach sufficiently high column
densities to turn molecular and allow star formation to occur. The
necessary column densities depend on metallicity: at $[{\rm Z}] \equiv
\log_{10}({\rm Z}/\Zsun) = -3.0$, gas must reach a column density of
$6.7 \times 10^3$ ($1.4 \times 10^4$, $8.0 \times 10^4$) $\msun$
pc$^{-2}$ in order to be 10 (50, 90) per cent molecular. At $[{\rm Z}]
= -2.5$ the required column is reduced to $2.3 \times 10^3$ ($4.9
\times 10^3$, $3.0 \times 10^4$) $\msun$ pc$^{-2}$, and it is only
$8.0 \times 10^2$ ($1.7 \times 10^3$, $1.0 \times 10^4$) $\msun$
pc$^{-2}$ at $[{\rm Z}] = -2.0$. This naturally explains the observed
dependence of the $\fstar$ suppression scale on the amplitude of the
metallicity floor. It implies that the true $\fstar$ suppression
realized in nature will depend on the details of the enrichment
history of a given halo, which should lead to a broadening of the
$\Mhalo$ dependence of the $\fstar$ suppression.

\section{Comparisons with high redshift observations}
\label{sec:evolution}

\subsection{Evolution of the Luminosity Function}

\begin{figure}[tp]
\includegraphics[width=\columnwidth]{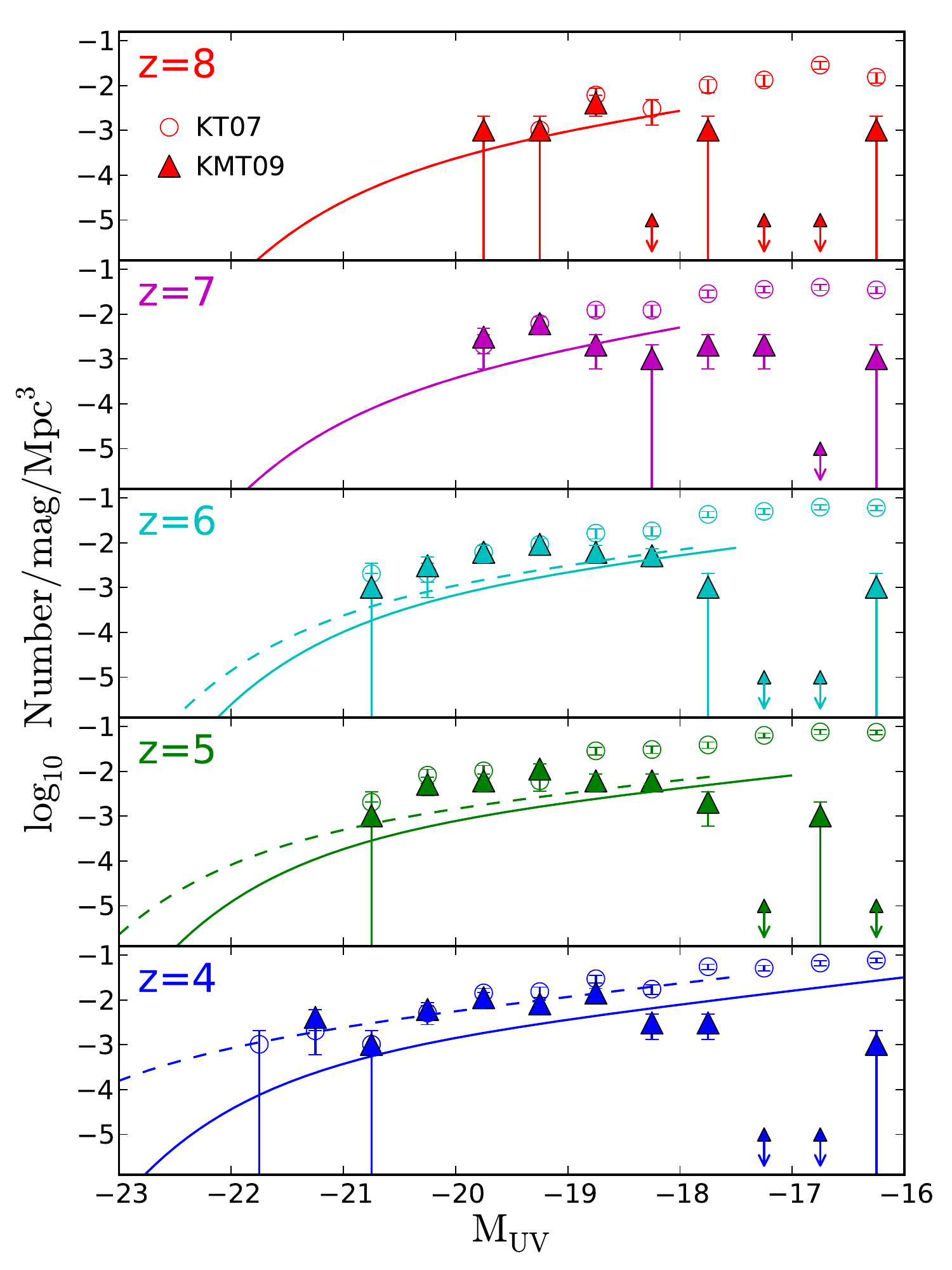}
\caption{The luminosity function in the KT07 (open circles) and KMT09\_ZFz10
  (solid triangles) simulations compared to observational results. We
  calculate UV luminosities from the simulated galaxies' SFR according
  to ${\rm L_{UV}} = 8.0 \times 10^{27} \, ({\rm SFR}/\msun\,{\rm
    yr}^{-1})$ erg s$^{-1}$ Hz$^{-1}$ (${\rm M_{UV}} = 51.63 - 2.5
  \log_{10}({\rm L_{UV}}/$erg s$^{-1}$ Hz$^{-1})$), corresponding to a
  Salpeter IMF from 0.1 to 125 $\msun$ \citep[same
    as][]{bouwens_ultraviolet_2011}. Error bars are statistical only. The solid
  lines are the uncorrected luminosity functions reported by
  \citet{bouwens_uv_2007,bouwens_ultraviolet_2011}, and the dashed lines the
  same relations corrected for dust extinction by (1.55, 0.625, 0.375,
  0, 0) magnitudes at $z=(4,5,6,7,8)$ \citep[cf. Table 8 in
  ][]{bouwens_ultraviolet_2011}.}
\label{fig:LF_vs_z}
\vspace*{0.1in}
\end{figure}

In Figure~\ref{fig:LF_vs_z} we present a comparison of the luminosity
functions (LF) from the KT07 and KMT09\_ZFz10 simulations and the recent high
redshift determinations from \citet{bouwens_uv_2007,bouwens_ultraviolet_2011}
based on deep HST ACS and WFC3 observations of $B$ ($\langle z \rangle
= 3.8$), $V$ (5.0), $i$ (5.9), $z$ (6.8), and $Y$-band (8.0) dropout
galaxies. Following \citet[hereafter B10]{bouwens_ultraviolet_2011}, we
determine UV luminosities from the simulated SFR according to ${\rm
  L_{UV}} = 8.0 \times 10^{27}$ (SFR / $\msun$ yr$^{-1}$) erg s$^{-1}$
Hz$^{-1}$, corresponding to a Salpeter IMF from 0.1 to 125 $\msun$ and
a constant SFR over $\gtrsim 100$ Myr \citep{madau_star_1998}. We
calculate SFR from our simulated galaxies by summing the mass of all
young star particles with ages less than $\tau_\star = 30$ Myr and
dividing by the SF time scale,
\begin{equation}
{\rm SFR} = \sum_{{\rm age}\,<\,\tau_\star} \f{m_\star}{\tau_\star}. \label{eq:SFR}
\end{equation}
The assumption of a constant SFR over the past 100 Myr is not likely
to hold for our simulated galaxies, and we therefore may be
overestimating their UV luminosities by a factor $\sim 2$ or so. A
Kroupa IMF, on the other hand, would result in $\sim 1.7$ higher UV
luminosity for a given SFR.  We compare the simulated LFs to the
Schechter function fits reported by B10 (shown in solid lines), as
well as to these same relations corrected for dust extinction (dashed
lines) by (1.55, 0.625, 0.375, 0, 0) magnitudes at $z=(4,5,6,7,8)$
(cf. Table 8 in B10).

The KT07 LF exceeds the uncorrected B10 LF by $\sim$ 1 dex at all
redshifts less than 8, and over the entire range of luminosities
probed by B10. Applying the B10 dust correction brings the $z=4$ LF
into agreement with the KT07 simulation. This agreement is remarkable,
since it results without any tuning of our models. It is puzzling,
however, since it seems to imply that there is no dwarf galaxy problem
at $z=4$. In general, the steep faint-end LF slopes ($-1.7$ to $-2.0$)
at $z \geq 4$ reported by B10 appear to be in tension with the need to
strongly suppressed star formation efficiency in low mass halos in
order to match the lower redshift stellar mass functions and dwarf
galaxy luminosity functions.  Regardless of this, at $z>4$ the
disagreement between our KT07 LFs and the dust-corrected observational
ones remains substantial.

The KMT09\_ZFz10 LF matches the KT07 one at high luminosities, reflecting the
fact that \HH-regulation is not effective in high mass (high ${\rm
  L_{UV}}$) halos. Close to the sensitivity limit of the B10
observations, the KMT09\_ZFz10 LF begins to roll over, improving the
agreement between simulations and observations. This rollover,
however, continues to lower luminosities, and it appears that in the
KMT09\_ZFz10 simulation the \HH-suppression may be too efficient at low UV
luminosities, at least if the faint end of the observed UV LF
continues to rise to lower luminosities. As we discussed in
\S~\ref{sec:fstar_resolution}, the exact mass scale of the SF
suppression (and hence the downturn of the LF) is sensitive to the
nature of the metal enrichment process, and also depends on the
resolution of our simulations.

\subsection{Evolution of the Stellar Mass Density}

\begin{figure}[tp]
\includegraphics[width=\columnwidth]{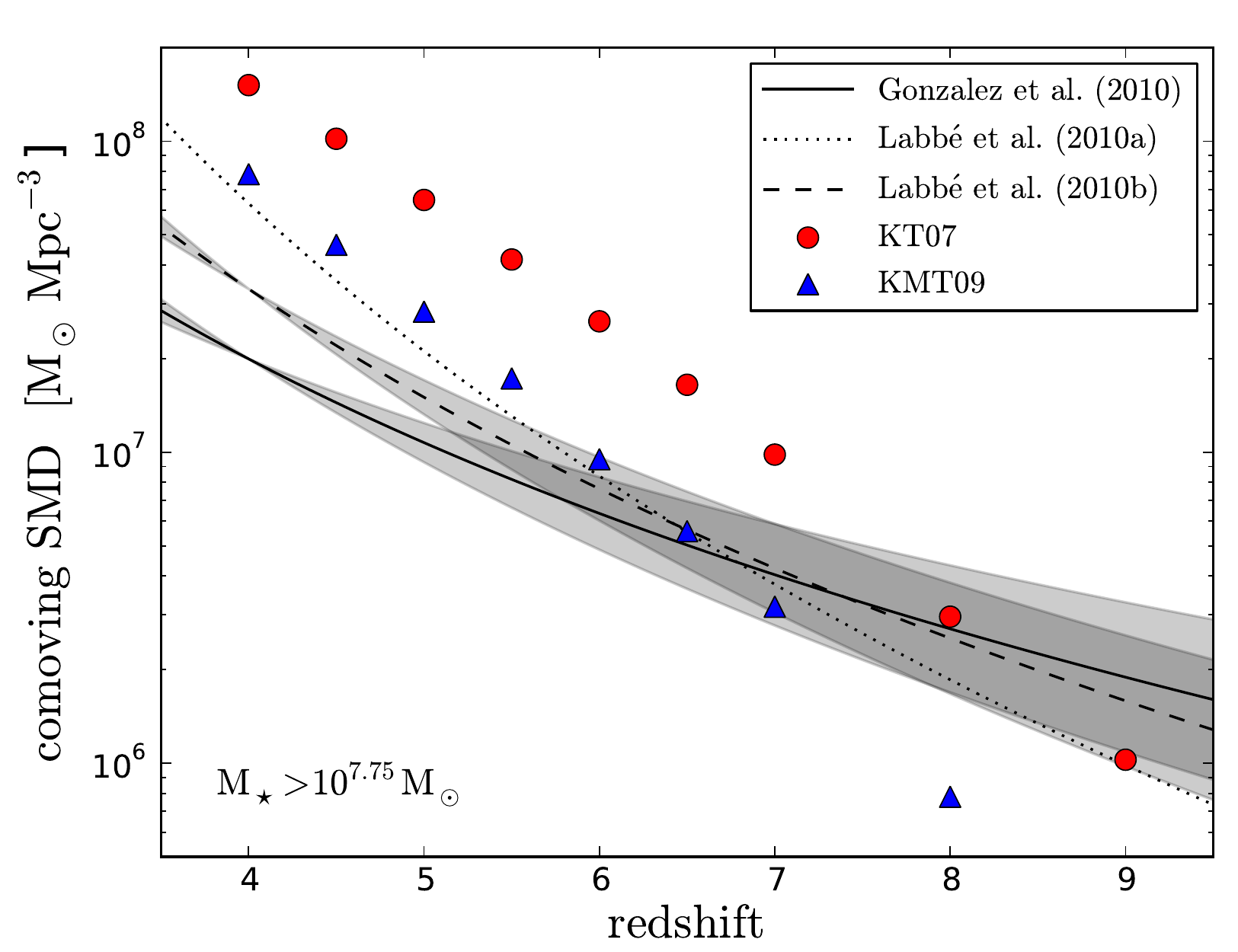}
\caption{The stellar mass density for galaxies with M$_\star>10^{7.75}
  \msun$ vs. z in the KT07 and KMT09\_ZFz10 simulations, compared to
  observational determinations from \citet{gonzalez_evolution_2011}
  (SMD $\propto(1+z)^{-3.4 \pm 0.8}$, solid line),
  \citet{labbe_ultradeep_2010}, (SMD $\propto (1+z)^{-6}$, dotted
  line), and \citet{labbe_xxx_star_2010} (SMD $\propto (1+z)^{-4.4 \pm
    0.7}$, dashed line). The uncertainty in the slope reported by
  \citet{gonzalez_evolution_2011} and \citet{labbe_xxx_star_2010} is
  indicated by the gray regions, arbitrarily normalized at $z=4$.}
\label{fig:rhostar_vs_z}
\vspace*{0.1in}
\end{figure}

In Figure~\ref{fig:rhostar_vs_z} we show the redshift evolution of the
co-moving stellar mass density (SMD) contributed by galaxies with
M$_\star > 10^{7.75} \msun$ in the KT07 and KMT09\_ZFz10 simulations,
compared to the observational determination from
\citet{gonzalez_evolution_2011}. These authors used restframe optical
photometry from Spitzer/IRAC to infer stellar masses for a sample of
$\sim 300$ $z=4$ galaxies, and combined these with HST ACS and WFC3/IR
restframe UV fluxes to establish an empirical stellar mass to UV
luminosity relation. Assuming this same M$_\star$-L$_{\rm UV}$ extends
to higher redshifts, they integrated the \citet{bouwens_ultraviolet_2011}
$z=4-7$ restframe UV luminosity function down to a limiting magnitude
of $M_{\rm UV}=-18$ (corresponding to M$_\star=10^{7.75} \msun$), and
obtained an empirical estimate of the SMD evolution: SMD $\propto
(1+z)^{-3.4 \pm 0.8}$, shown as the solid line and gray region in
Figure~\ref{fig:rhostar_vs_z}.

To demonstrate the systematic uncertainty in this kind of empirical
determination of the stellar density evolution, we also show results
from \citet{labbe_ultradeep_2010} (dotted line, SMD $\propto
(1+z)^{-6}$) and \citet{labbe_xxx_star_2010} (dashed line, SMD $\propto (1+z)^{-4.4 \pm 0.7}$), who performed similar work including
$z=8$ Y-band dropouts, but used M$_\star$-L$_{\rm UV}$ relations
derived from different low redshift data
\citep{stark_evolutionary_2009}.

Likewise our simulated SMD's are not free from systematic
uncertainties: on the one hand they are likely somewhat
underestimated, especially at high redshift, since the simulations'
limited box size (12.5 Mpc) results in a delayed formation of massive
halos, owing to the absence of density perturbations in the initial
conditions with wavelengths exceeding the box size
\citep{tormen_adding_1996}. On the other hand the lack of effective
stellar feedback in our simulations probably artificially enhances the
stellar mass density at all redshifts.

Given the substantial uncertainties in both observations and
simulations, we consider the agreement between the two to be
satisfactory. We do note, however, that the simulations prefer a
somewhat steeper evolution of the SMD than
\citet{gonzalez_evolution_2011}, and that the \HH-regulated SF reduces
the SMD by a factor $\sim 3-5$ over the covered redshift range
compared to the standard SF case. The difference between KT07 and
KMT09\_ZFz10 decreases towards lower redshifts, which can be attributed to
the increase of the typical halo mass (the knee in the Press-Schechter
massfunction) with decreasing redshift, resulting in an increasing
fraction of resolved halos with mass above the $\fstar$-suppression
mass scale.

\subsection{Evolution of the Star Formation Rate Density}

\begin{figure}[tp]
\includegraphics[width=\columnwidth]{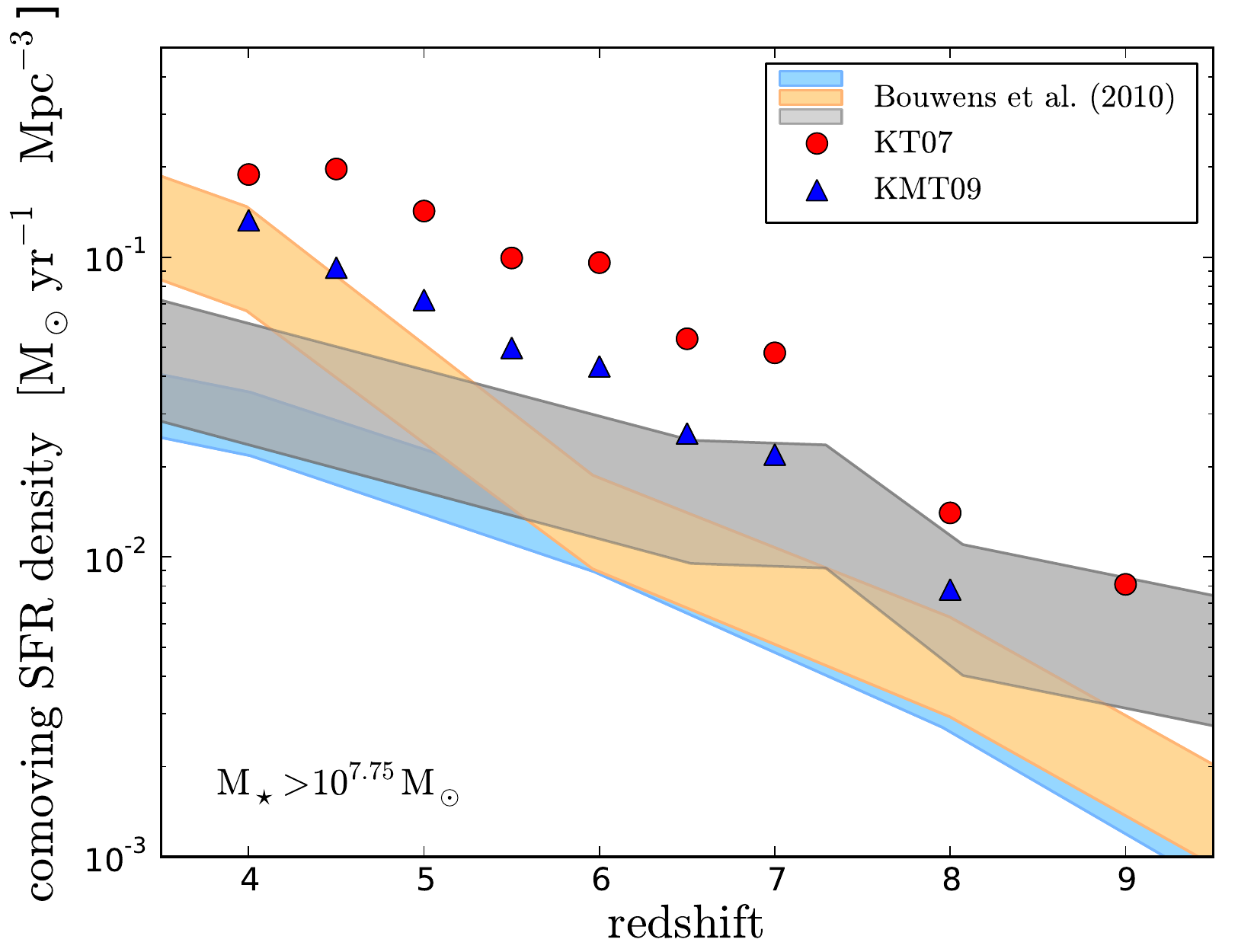}
\caption{The SFR density for galaxies with $M_\star > 10^{7.75} \msun$
  vs. z in the KT07 and KMT09\_ZFz10 simulations, compared to observational
  determinations from \citet{bouwens_ultraviolet_2011}, which were obtained by
  converting restframe UV luminosities to SFR \citep{madau_star_1998},
  without (blue band) and with dust corrections (orange band), and by
  converting stellar mass densities
  \citep{stark_evolutionary_2009,labbe_ultradeep_2010,labbe_xxx_star_2010,gonzalez_evolution_2011}
  to SFR densities (light gray band).}
\label{fig:rhoSFR_vs_z}
\vspace*{0.1in}
\end{figure}

In Figure~\ref{fig:rhoSFR_vs_z} we extend the comparison between our
simulations and high-z observational data to the evolution of the SFR
density, obtained by summing the SFR (Eq.~\ref{eq:SFR}) of all
galaxies with M$_\star > 10^{7.75} \msun$ (${\rm M_{UV}} < -18$) and
dividing by the box volume. The observational data again comes from
B10, who derived SFR densities by converting their restframe UV
luminosities to SFR according to \citet{madau_star_1998}, and
integrated down to a limiting magnitude of $M_{\rm UV}=-18$, with
(orange band) and without (blue band) a dust correction at $z \lesssim
6$ from \citet{bouwens_uv_2009}. Note that the very blue UV continuum
slopes observed in $z \gtrsim 7$ galaxies
\citep{bouwens_very_2010,bunker_contribution_2010,finkelstein_stellar_2010,oesch_z_2010}
imply low dust abundances and correspondingly weak extinction
corrections at these high redshifts. For comparison we also show
(light gray band) the SFR density implied by the SMD evolution
discussed above, assuming a fixed SMD $\propto (1+z)^{-4.4}$
extrapolation to $z \gtrsim 8$ \citep{labbe_xxx_star_2010}.

Our simulated SFR densities are somewhat higher than those reported by
\citet{bouwens_ultraviolet_2011}. \HH-regulated SF, however, reduces the SFR
densities by about a factor of two, and brings them close to agreement
with those determined from the SMD evolution, and only a factor $\sim
3$ above the luminosity density derived values at $z>6$. At lower
redshifts, the rise due to dust corrections in the observational SFR
densities reduces the difference, and an extrapolation of the KMT09\_ZFz10
SFR densities to $z=4$ is in good agreement with the dust corrected
observational determination. If the observed steep UV continuum slopes
are a selection effect and not representative of typical $z \gtrsim 7$
galaxies, then a more heavily dust obscured population of galaxies
that has been missed by current surveys could close the high redshift
gap between our simulations and current data. Additionally it is
possible that a more effective stellar feedback implementation in our
simulations could further reduce the simulated SFR densities.

\section{Summary and Discussion}
\label{sec:conclusions}

Motivated by the observational fact that star formation, both in the
local universe \citep[e.g.][]{bigiel_star_2008} and at intermediate
redshifts \citep[e.g.][]{genzel_study_2010}, correlates more tightly
with the density of gas in the molecular phase than the total gas
density, we have implemented an \HH-regulated SF prescription in
cosmological galaxy formation simulations, with the goal of
investigating whether this new ingredient can help to alleviate
outstanding problems in our theoretical understanding of dwarf galaxy
formation.

Rather than following the non-equilibrium molecular hydrogen chemistry
including time-dependent and spatially inhomogeneous radiation transfer
of the ionizing and \HH-dissociating stellar radiation field (G09), we
utilize in our simulations the results of one-dimensional radiative
transfer calculations of the \HH\ formation-dissociation balance in an
idealized spherical giant atomic--molecular complex subject to a
uniform and isotropic Lyman-Werner (LW) radiation field
\citep{krumholz_atomic--molecular_2008,krumholz_atomic--molecular_2009,mckee_atomic--molecular_2010}. These
calculations showed that under the assumption of two-phase equilibrium
in the ISM, the \HH\ abundance is determined entirely by the column
density and metallicity of gas on $\sim 100$ pc scales -- comparable
to the cell size of our highest resolution grid cells.

We have deliberately chosen a weak implementation of supernova
feedback in order to isolate the effects of the molecular hydrogen
chemistry on star formation in a cosmological setting. Remarkably, we
find that accounting for the transition from HI to \HH\ alone is able
to fulfill one of the functions of the neglected supernova feedback,
namely the suppression of star formation in low mass halos. The main
results of our study are summarized as follows.

\begin{enumerate}
\item Both our conventional and \HH-regulated SF prescriptions are
  able to reproduce the observational scaling relation between the SFR
  surface density and the total gas density, the Kennicutt-Schmidt
  relation (Fig.~\ref{fig:Kennicutt}).

\item With the conventional SF prescription the observed turnover of
  $\SigmaSFR$ at low $\SigmaGas$ is recovered by manually tuning a SF
  threshold density. The \HH-regulated prescription, however,
  automatically reproduces this cutoff
  (Fig.~\ref{fig:Kennicutt_nthresh}) and thus reduces the number of
  free parameters in the SF prescription by one.

\item We are able to reproduce many of the observational results
  pertaining to the KS relation, for example the slope of the
  molecular gas KS relation ($\SigmaSFR-\Sigma_{\rm H_2}$) and the
  much weaker correlation of $\SigmaSFR$ with the atomic gas surface
  density (Fig.~\ref{fig:Kennicutt_H2}). We also recover the observed
  metallicity dependence of the low $\SigmaGas$ turnover in the KS
  relation (Fig.~\ref{fig:Kennicutt_Z}), which occurs at higher
  $\SigmaGas$ in lower metallicity systems, reflecting the
  atomic-to-molecular gas transition. Lastly, in agreement with recent
    observations, we see an increased scatter in the \HH\ KS relation
    for smaller spatial smoothing scales.

\item We expect an increased scatter in the KS relation at higher
  redshifts, since the typically larger stellar velocity dispersions
  in high redshift systems will allow young stars to wander outside of
  high column density regions.

\item \HH-regulation suppresses SF in low mass halos, reducing the
  need for stellar feedback (Fig.~\ref{fig:fstar}), as previously
  reported by G09 and GK10 for a set of cosmological zoom-in galaxy
  formation simulations including non-equilibrium \HH\ formation and
  radiative transfer. We confirm these prior results with better
  statistics (hundreds of halos) and without the need for a
  complicated and expensive radiative transfer treatment.

\item The halo mass dependence appears to be tied primarily to a
  difference in metal enrichment, rather than gas column density
  (Fig.~\ref{fig:SigmaZphase}). On the other hand, for a given
  metallicity floor (set for example by the first generation of
  Pop.III supernovae), the star forming ability of halos is determined
  by the highest column densities their gas can condense to. Low mass
  halos don't have sufficiently high column density gas to allow the
  transition to molecular phase and hence star formation.

\item Suppressing the star formation efficiency in low mass halos
  lowers the cosmic stellar mass density and star formation rate
  density. We find reasonable agreement between our \HH-regulated
  simulation and observational determinations of the evolution of the
  SMD and SFR-density at $z>4$. Both simulations and observations are
  subject to large systematic uncertainties.

\end{enumerate}

As we discussed throughout the text, a number of caveats apply to our
findings. First, the results we have presented here are not completely
independent of the implementation details of our simulations
(e.g. 2-phase equilibrium assumption, metallicity floor, numerical
resolution). We therefore cannot claim with any confidence to have
pinpointed the halo mass below which star formation is suppressed due
to the inability of gas to become molecular.

Second, given our computational resources, we were unable to continue our simulations beyond $z=4$. It is possible that a large build-up of atomic gas will lead to a burst of star formation at lower redshift ($z \simeq 1-3$), when metallicities have increased sufficiently to allow the transition to \HH\ to occur, and in fact this may help to explain how the specific SFR of $10^{10} - 10^{12} \msun$ halos can exceed the instantaneous gas accretion rate at these redshifts \citep{krumholz_metallicity-dependent_2011}.

Lastly, like most cosmological galaxy formation simulations to date, our simulated galaxies suffer from the so-called ``baryonic overcooling problem'', resulting in unrealistically high central densities (and hence strongly peaked circular velocity curves) and stellar mass fractions in our high mass halos that are too large compared to observations. Although we cannot exclude the possibility that these artificially high central densities affect the results of our study, we believe that this makes our results conservative, in the sense that a more realistic simulation with lower central gas densities would likely experience even more \HH-regulated SF suppression. Reduced central densities would demand higher metallicity for the transition to \HH\ to occur, and at the same time they would allow the metals ejected by stars to become diluted more easily, thereby reducing the metallicity of the high density star-forming gas.


Even allowing for these caveats, we believe that the
atomic-to-molecular gas transition may play an important role in
regulating star formation in low mass halos. It may also help to
explain the result recently reported by
\citet{boylan-kolchin_too_2011} that collisionless simulations of
galactic dark matter substructure (Via Lactea II and Aquarius) predict
a number of subhalos too centrally concentrated to host any of the
known Milky Way dwarf satellite galaxies. As
\citet{boylan-kolchin_too_2011} suggest, one possible explanation of
this puzzle is that star formation in dark matter halos becomes
stochastic below some mass. The sensitivity of an \HH-regulated SF
prescription to early metal enrichment may provide the necessary
stochasticity.

In our simulations, \HH-regulation quenches SF in dwarf galaxies at
the outset, without the need to artificially enhance stellar feedback
by turning off gas cooling or hydrodynamically decoupling
momentum-driven winds, as is commonly done in the literature. These
latter implementations of enhanced stellar feedback quench SF in dwarf
galaxies due to the inability of their shallow potential wells to
retain the resulting gas outflows. In \HH-regulated SF, by contrast,
it is the inability of low mass halos to accumulate a sufficiently
large column density of metal-enriched material that results in
suppressed SF. Efficient supernova feedback implies the presence of a
large pool of hot gas as well as galactic winds, neither of which are
necessarily present in an \HH-regulated scenario. This difference may
be be a good way to observationally distinguish the two mechanisms.

Taken at face value, our results imply that many low mass dark matter
halos at high redshift should be filled with relatively cold, yet
atomic gas, which is prevented from becoming molecular by its low
metallicity. While neutral gas surveys with radio telescopes are hard
pressed to reach beyond the local universe, constraints on the HI mass
function do exist at low redshift. The ALFALFA survey
\citep{martin_arecibo_2010}, for example, has reported a $z \lesssim
0.06$ HI mass function down to $\log(M_{\rm HI}/\msun) = 6.2$, that is
well fit by a Schechter function with a low mass slope of
$\alpha=-1.33$. This is considerably shallower than the mass function
of dark matter halos \citep[$\alpha \approx
  -1.8$,][]{boylan-kolchin_resolving_2009}, indicating that not all
low mass dark matter halos can be allowed to retain their full
baryonic content in the form of cold, atomic gas down to the present
epoch. Another constraint may come from the statistics of damped
Lyman-$\alpha$ (DLA) systems at $z \lesssim 5$
\citep{prochaska_nonevolution_2009,noterdaeme_evolution_2009}.
Whether the observed frequency distribution of DLA column densities
and their covering fraction is consistent with large amounts of atomic
gas in high redshift dwarf galaxies should be investigated in detail.

Strongly suppressing star formation in low mass halos at high redshift could make it difficult for faint galaxies to reionize the universe by $z=6$, as is commonly advocated \citep[e.g.][]{madau_radiative_1999}. Indeed, in our simulations the total number of hydrogen ionizing photons per hydrogen atom\footnote{We assume 4000 hydrogen ionizing photons per stellar baryon for every solar mass of stars formed, as appropriate for a Salpeter IMF from 0.1 to 125 $\msun$ and a mean ionizing photon energy of 20 eV \citep{madau_radiative_1999,leitherer_starburst99:_1999}.} produced by $z=6$ is reduced from $\sim 18$ in KT07 to $\sim 6$ in KMT09. As shown by \citep{kuhlen_concordance_2012}, a suppression scale of $\Mhalo = 10^{10} \msun$ can simultaneously satisfy reionization and lower redshift Lyman-$\alpha$ forest constraints only if the escape fraction of ionizing radiation evolves strongly from $z=4$ to higher redshifts. A slightly smaller suppression mass of $\sim 10^9 \msun$ \citep[e.g. as in][]{krumholz_metallicity-dependent_2011}, however, is in good agreement with all constraints.

While we have emphasized in this work that \HH-regulated SF can
perform some of the functions typically assigned to supernova
feedback, we of course acknowledge that supernovae do in fact occur in
nature and that their associated injection of energy and momentum into
the surrounding interstellar medium is likely to significantly impact
subsequent star formation, and may help to solve the problem of
forming galaxies that are too concentrated in the simulations. Future
research will be necessary to elucidate how the interplay of molecular
chemistry and supernova feedback shapes star formation in dwarf
galaxies and beyond.

\acknowledgments

MK acknowledges stimulating and fruitful discussions with F.~Bigiel,
C.-A.~Faucher-Gigu\`{e}re, R.~Feldmann, N.~Gnedin, P.~Hopkins,
D.~Kere\v{s}, A.~Kravtsov, M.~McQuinn, N.~Murray, M.~Norman, S.~Skory,
and E.~Quataert. We thank A.~Bolatto for providing us with a table of
\HH\ and HI column densities towards the SMC. MRK acknowledges support
from: an Alfred P. Sloan Fellowship; the NSF through grants
AST-0907739 and CAREER-0955300; and NASA through Astrophysics Theory
and Fundamental Physics grant NNX09AK31G and through a Spitzer Space
Telescope Theoretical Research Program grant. PM acknowledges support
from NASA through grant NNX09AJ34G and from NSF through grant
AST-0908910. JHW is supported by NASA through Hubble Fellowship grant
\#120-6370 awarded by the Space Telescope Science Institute, which is
operated by the Association of Universities for Research in Astronomy,
Inc., for NASA, under contract NAS 5-26555.

A special thank you to all Enzo developers for being such a great
community and contributing your time and expertise to the development
and improvement of the Enzo code. Analysis of the simulations was
greatly aided by the yt software \citep{turk_yt:_2011}.


\bibliographystyle{apj}
\bibliography{MyZoteroLibrary}

\begin{appendix}

\section{Sobolev-like approximation of $\Sigma$}

\begin{figure}[htp]
\includegraphics[width=0.49\textwidth]{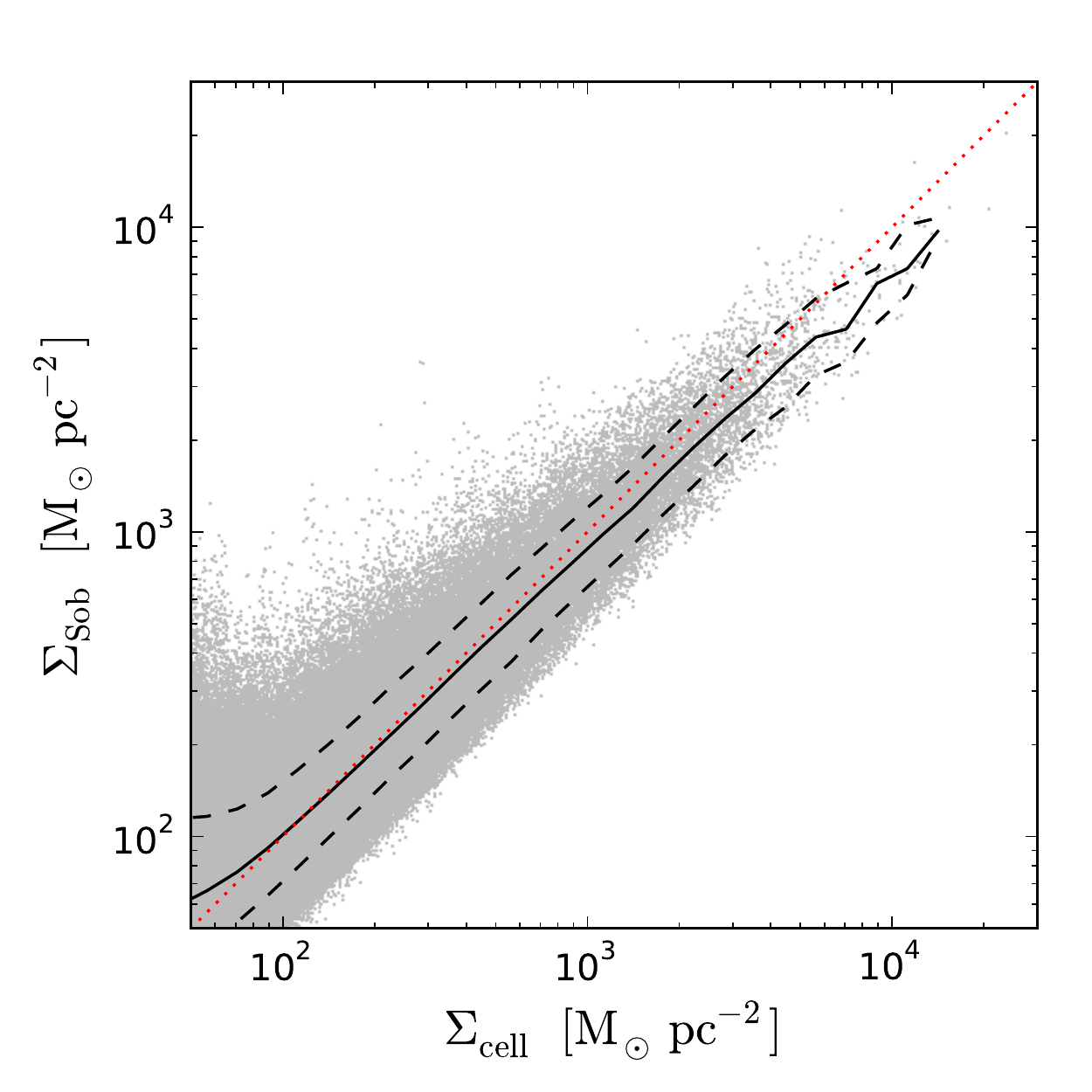}
\includegraphics[width=0.465\textwidth]{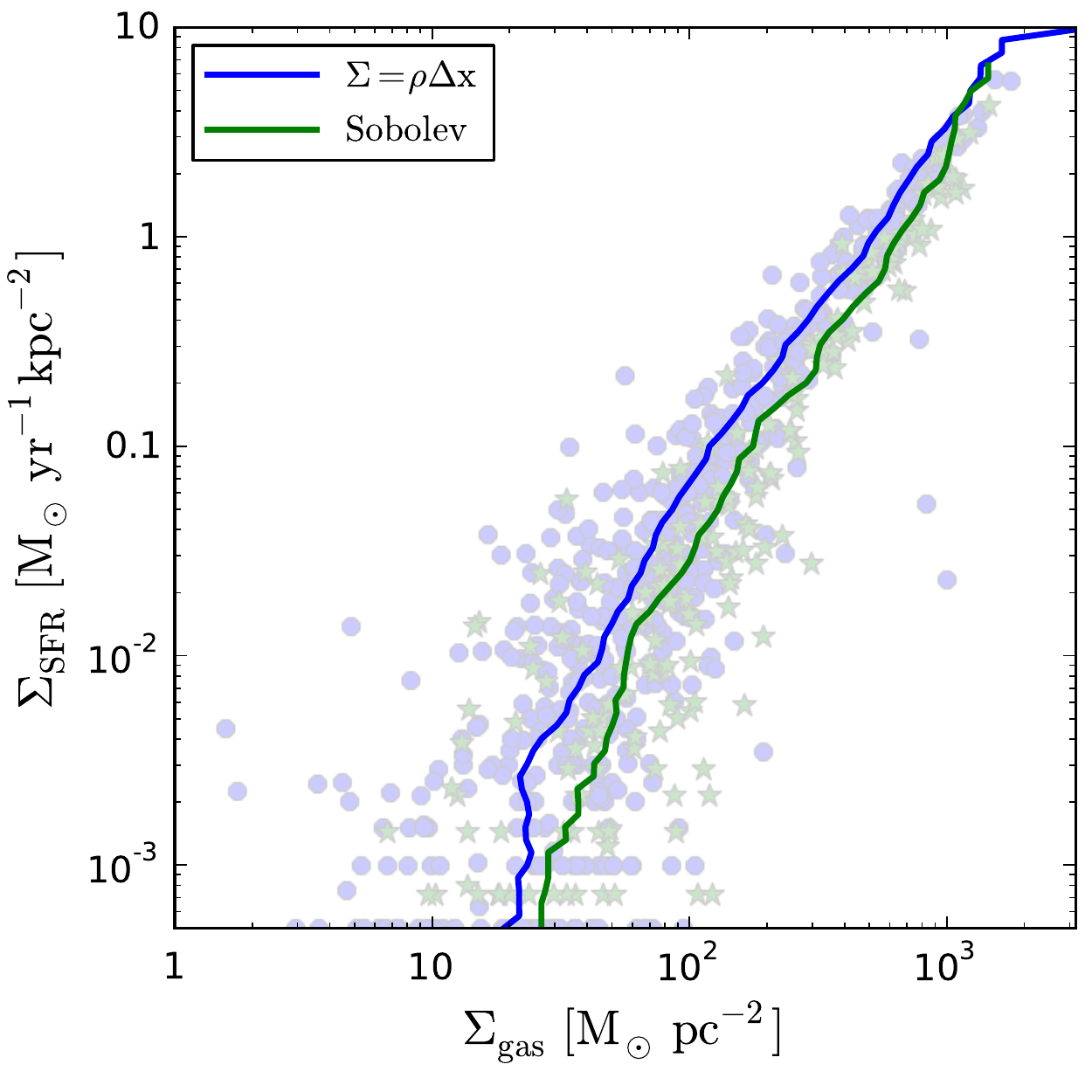}
\caption{\textit{Left:} Comparison of a Sobolev-like estimate of the
  column density ($\SigmaSob = \rho \times (\rho / \nabla \rho)$, see
  text for details) with the simple grid-cell column ($\Sigma_{\rm
    cell} = \rho \Delta x$), obtained by post-processing the $z=4$
  KMT09 output. Only columns greater than 50 M$_\odot$ pc$^{-2}$ are
  shown, since the \HH\ content is negligible at lower columns. The
  solid and dashed black lines show the median and 16$^{\rm th}$ -
  84$^{\rm th}$ percentiles of $\Sigma_{\rm Sob}$ for a given
  $\Sigma_{\rm cell}$. The Sobolev-like estimate results in slightly
  lower columns, especially at high columns. \textit{Right:} KS
  relation for simulation KMT09\_Sob, in which the Sobolev-like
  approximation was used at simulation run-time to calculate the
  column density entering the KMT09 expression for $\fHH$. Note that
  we did not use the Sobolev-like approximation to determine the
  x-axis quantity $\Sigma_{\rm gas}$, the column smoothed at $l=3$
  ($\sim$ 1 kpc). The slightly lower columns with the Sobolev-like
  approximation result in lower $\fHH$ and hence a slightly reduced
  SFR at a given $\Sigma_{\rm gas}$.}
\label{fig:Sobolev}
\end{figure}

One way to alleviate the explicit resolution dependence in our
implementation of the KMT09 algorithm
(Eq.~\ref{eq:KMT09model_begin}-\ref{eq:KMT09model_end}) is to use a
Sobolev-like approximation of the column density
\citep[see][]{gnedin_modeling_2009}, instead of simply multiplying the
grid cell density by its width. We have implemented a Sobolev-like
approximation as follows:
\begin{center}
\begin{minipage}{2.5in}
\begin{eqnarray}
\rho_x^+ & = & 0.5 \, (\rho_{i,j,k} + \rho_{i+1,j,k}), \nonumber \\
h_x^+ & = & \frac{ \rho_x^+ }{ |\rho_{i+1,j,k} - \rho_{i,j,k}| / \Delta x }, \nonumber \\
\Sigma_x^+ & = & \rho_x^+ \cdot h_x^+, \nonumber
\end{eqnarray}
\end{minipage}
\begin{minipage}{2.5in}
\begin{eqnarray}
\rho_x^- & = & 0.5 \, (\rho_{i,j,k} + \rho_{i-1,j,k}), \nonumber \\
h_x^- & = & \frac{ \rho_x^- }{ |\rho_{i,j,k} - \rho_{i-1,j,k}| / \Delta x }, \nonumber \\
\Sigma_x^- & = & \rho_x^- \cdot h_x^-,
\end{eqnarray}
\end{minipage} \\
\end{center}
and similarly for the y and z directions. This definition has the
virtue that it is resolution-independent, at least to the extent that
the density field itself is resolution-independent. To obtain a total
column density, we take the harmonic mean over the six cardinal
directions,
\begin{equation}
\langle \Sigma \rangle = \frac{6}{1/\Sigma_x^+ + 1/\Sigma_x^- + 1/\Sigma_y^+ + 1/\Sigma_y^- + 1/\Sigma_z^+ + 1/\Sigma_z^-}.
\end{equation}

We have applied this Sobolev-like approximation for $\Sigma_{\rm gas}$
in a post-processing analysis of the $z=4$ output of the KMT09
simulation, and compared the resulting column density with the simple
grid cell based estimate used in the simulation, see the left panel of
Fig.~\ref{fig:Sobolev}. We're only showing points with $\Sigma > 50$
M$_\odot$ pc$^{-2}$, since at lower columns the \HH\ abundance, and
hence SFR, is negligible. The median of $\Sigma_{\rm Sob}$ lies close
to the 1-1 line, but drops below it around $\Sigma_{\rm cell} = 100$
M$_\odot$ pc$^{-2}$, implying that the Sobolev-like approximation
systematically yields slightly lower values at high $\Sigma_{\rm
  cell}$. At $\Sigma_{\rm cell} =$ 100, 1000, and 5000 M$_\odot$ pc$^{-2}$
the median of $\Sigma_{\rm Sob}$ is 101, 880, and 3930 M$_\odot$
pc$^{-2}$, respectively. The 1-$\sigma$ scatter of $\Sigma_{\rm Sob}$
around the median is 0.26 dex (i.e. 68\% of all cells lie within a
factor of $\sim 1.8$ of the median).

\begin{figure}[htp]
\centering
\includegraphics[width=0.465\textwidth]{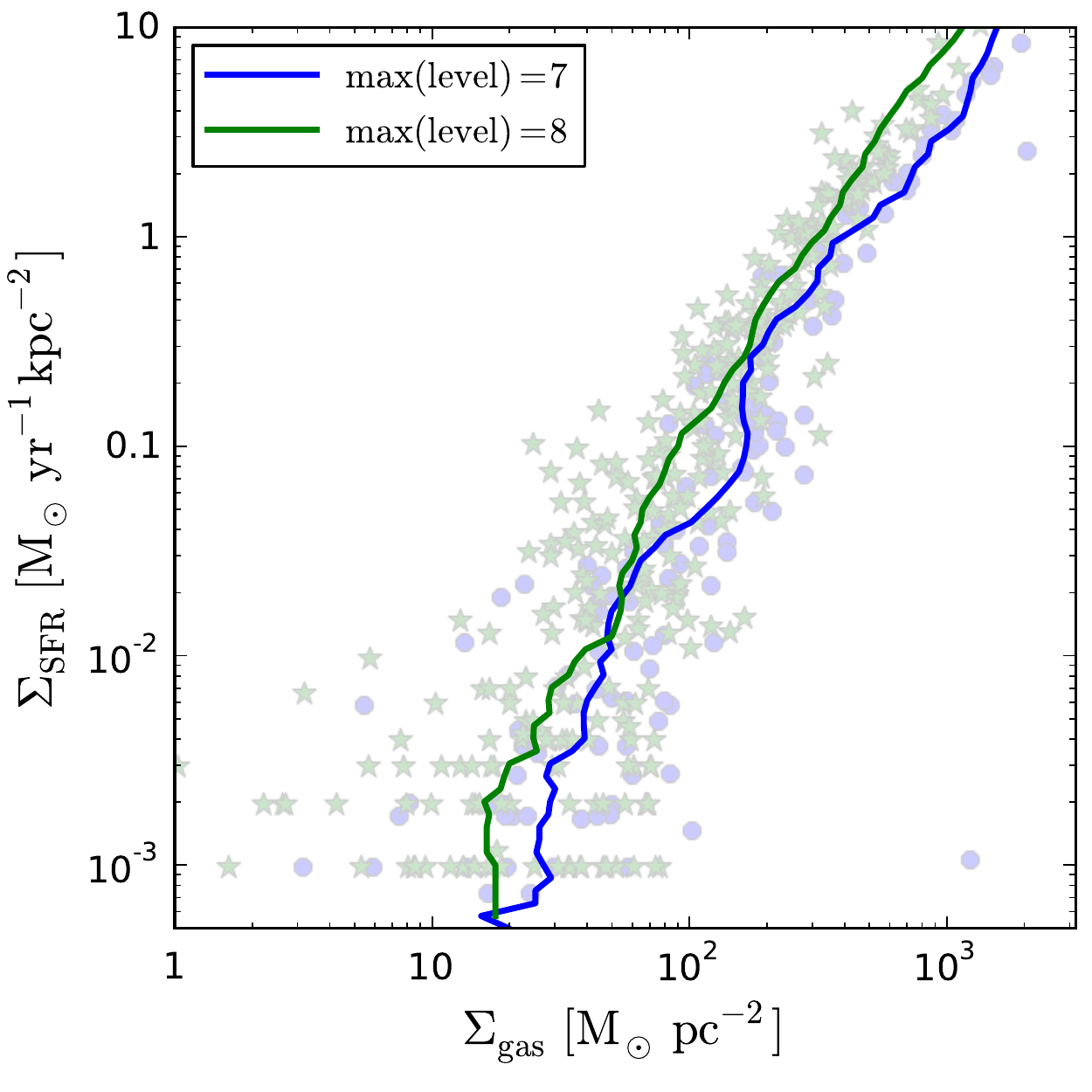}
\caption{Comparison of the KS relation for the KMT09\_Sob
  and KMT09\_SobL8 simulations at $z=6$. The use of $\SigmaSob$ does
  not fully remove the resolution dependence, but reduces it mildly
  compared to simulations with $\SigmaCell$ (cf. right panel of
  Fig.~\ref{fig:KS_resolution}).}
\label{fig:Sobolev_L8}
\end{figure}

This post-processing analysis indicates that a simulation actually
using the Sobolev-like approximation at run-time might have somewhat
reduced SF, owing to the systematically slightly lower columns, which
provide less shielding and hence lower \HH\ abundances. This is in
fact borne out in practice, as shown in the right panel of
Fig.~\ref{fig:Sobolev}. We ran two additional simulations, KMT09\_Sob and
KMT09\_SobL8, in which the column densities entering the KMT09 $\fHH$
prescription were determined at run-time using the Sobolev-like
estimator. Indeed the resulting KS relation has a slightly lower
normalization, indicating a somewhat reduced SF efficiency.

As discussed in the main text, to some degree there will always be a
resolution dependence in the determination of the column density in
our simulations, simply because we are not resolving the true Jeans
length and are applying a minimum pressure support in order to prevent
artificial fragmentation. This is demonstrated in
Fig.~\ref{fig:Sobolev_L8}, which shows the KS relations for the
KMT09\_Sob and KMT09\_SobL8 simulations. The corresponding plot for
simulations with $\SigmaCell$ is the right panel of
Fig.~\ref{fig:KS_resolution}, and a comparison between these two
figures shows that the use of $\SigmaSob$ only mildly reduces the
resolution dependence.

The two different methods of estimating column densities ($\SigmaCell,
\SigmaSob$) are resolution-dependent in different ways. The difference
between the two in $\Sigma$ and in the KS relation is fairly small
(less than a factor of 2), and does not impact the overall conclusion
of our study that the metallicity-dependent nature of the
atomic-to-molecular transition can play a major role in explaining the
low star formation efficiency in dwarf galaxies.

\end{appendix}

\end{document}